\documentclass[10pt,letterpaper]{article}
\usepackage[top=0.85in,left=1.1in,footskip=0.75in,marginparwidth=2in]{geometry}
\usepackage{amsmath}

\usepackage{algorithm,algorithmic}
\usepackage{amsfonts}
\usepackage{booktabs}
\usepackage{amssymb}

\usepackage[utf8]{inputenc}

\usepackage[numbers]{natbib}

\usepackage{float}

\usepackage{nameref,hyperref}

\usepackage{microtype}
\DisableLigatures[f]{encoding = *, family = * }

\raggedright
\usepackage{changepage}

\usepackage[aboveskip=1pt,labelfont=bf,labelsep=period,singlelinecheck=off]{caption}

\makeatletter
\renewcommand{\@biblabel}[1]{\quad#1.}
\makeatother

\usepackage{lastpage,fancyhdr,graphicx}
\usepackage{epstopdf}
\fancyheadoffset[L]{2.25in}
\fancyfootoffset[L]{2.25in}

\usepackage{color}

\definecolor{Gray}{gray}{.25}

\usepackage{graphicx}

\usepackage{sidecap}

\usepackage{wrapfig}
\usepackage[pscoord]{eso-pic}
\usepackage[fulladjust]{marginnote}
\reversemarginpar

\begin{document}

\begin{flushleft}
{\Large
\textbf\newline{\textbf{Interpreting artificial neural networks to detect genome-wide association signals for complex traits}}
}
\newline
\\
\textbf{Burak Yelmen\textsuperscript{1,*},
Maris Alver\textsuperscript{1},
Merve Nur Güler\textsuperscript{1},
Estonian Biobank Research Team\textsuperscript{1},
Flora Jay\textsuperscript{2},
Lili Milani\textsuperscript{1}}
\\
\bigskip
{1} Estonian Genome Centre, Institute of Genomics, University of Tartu, Tartu, Estonia
\\
{2} CNRS, INRIA, LISN, Paris-Saclay University, Orsay, France
\\

\bigskip
*burak.yelmen@ut.ee

\end{flushleft}

\section*{Abstract}
Investigating the genetic architecture of complex diseases is challenging due to the multifactorial and interactive landscape of genomic and environmental influences. Although genome-wide association studies (GWAS) have identified thousands of variants for multiple complex traits, conventional statistical approaches can be limited by simplified assumptions such as linearity and lack of epistasis in models. In this work, we trained artificial neural networks to predict complex traits using both simulated and real genotype-phenotype datasets. We extracted feature importance scores via different post hoc interpretability methods to identify potentially associated loci (PAL) for the target phenotype and devised an approach for obtaining p-values for the detected PAL. Simulations with various parameters demonstrated that associated loci can be detected with good precision using strict selection criteria. By applying our approach to the schizophrenia cohort in the Estonian Biobank, we detected multiple loci associated with this highly polygenic and heritable disorder. There was significant concordance between PAL and loci previously associated with schizophrenia and bipolar disorder, with enrichment analyses of genes within the identified PAL predominantly highlighting terms related to brain morphology and function. With advancements in model optimization and uncertainty quantification, artificial neural networks have the potential to enhance the identification of genomic loci associated with complex diseases, offering a more comprehensive approach for GWAS and serving as initial screening tools for subsequent functional studies.

\paragraph{Keywords:} Deep learning, interpretability, genome-wide association studies, complex traits, schizophrenia

\newpage

\section*{Introduction}
Understanding the genetic component of complex diseases is challenging due to high polygenicity and influence of confounding factors. Genome-wide association studies (GWAS) have been crucial for the discovery of thousands of disease-related genomic loci in the past two decades \citep{abdellaoui_15_2023}, and they continue to be relevant due to ever-expanding biobanks with extensive phenotype and genotype data, which enable the detection of the most minuscule but statistically significant signals. Despite the immense success and broad applications across various complex traits, conventional GWAS methodology still relies on linear models (i.e., logistic and linear regression) for detecting signals. More recently, linear mixed models have been utilised to incorporate random and fixed effects, primarily to account for relatedness and population structure without data pruning \citep{jiang_generalized_2021,mbatchou_computationally_2021}. While these advancements have generally improved GWAS outcomes, they do not diverge from the core assumptions of linearity and additivity in genetic models for complex traits. A natural alternative for nonlinear modelling of such traits would be machine learning, and particularly deep learning, since sufficiently large neural networks can approximate any bounded continuous function as defined in the universal approximation theorem \citep{lu_universal_2020}. Hence, not surprisingly, various models have been developed for the prediction of disease status from genetic data following the success of deep learning applications in many domains \citep{alharbi_review_2022, zhou_deep_2023}. Despite the state-of-the-art performance in prediction, a main drawback of employing such complex architectures is the inherent difficulty of interpretability due to the high number of parameters at various abstraction levels. In this context, the development and integration of interpretability methodology is not only important for understanding model biases and improving architectures, but also for detecting novel features (e.g., genomic loci, haplotypes, and specific genetic variants) associated with the target phenotype. Various approaches have been proposed to tackle interpretability for the discovery of trait-associated loci, ranging from intrinsically interpretable architectures based on domain knowledge, such as functional gene annotations \citep{van_hilten_gennet_2021, xu_interpretable_2022, liu_explainable_2022}, to post hoc analysis of trained models \citep{badre_explainable_2023,liu_phenotype_2019, mieth_deepcombi_2021, kassani_deep_2022, sharma_deep_2020}. Despite these important advancements, multiple challenges remain for the widespread use of interpretable deep learning in genomics. For instance, many of the proposed methods rely on either specific architectures, specific interpretability methods, or a priori biological knowledge. In this work, we \textbf{(i)} propose a general framework for identifying potentially associated loci (PAL) from neural network models trained for phenotype prediction using genome-wide SNP array data, \textbf{(ii)} introduce an approach for obtaining p-values for the detected PAL, \textbf{(iii)} compare the performance of various post hoc and model-agnostic interpretability methods using simulations, and \textbf{(iv)} apply our approach to the schizophrenia (SCZ) cohort at Estonian Biobank (EstBB) \citep{leitsalu_cohort_2015, milani_biobanking_2024}  demonstrating its utility for GWAS with potentially novel loci not reported in literature.

\section*{Materials and Methods}
\subsection*{Data}
We assembled the SCZ cohort from the EstBB dataset so that all patients (ICD codes F20-F29) would have at least one antipsychotic prescription and have their first diagnosis after 2006 with the diagnosis age between 14 and 60 \citep{alver_genetic_2024}. As controls for the SCZ cases, we first assembled a set of individuals without any mental, behavioural and neurodevelopmental disorders (no ICD F* diagnosis) and no record of antipsychotic prescriptions. From this filtered set, controls were selected to match the age, sex and BMI distribution of the case cohort with four controls per case using MatchIt \citep{ho_matchit_2011}. We used exact matching (method = “exact”) for sex and nearest neighbour matching (method = “nearest”) for BMI and age, resulting in 1814 cases (642 male, 1172 female) and 7325 controls (2586 male, 4739 female). In addition, the whole dataset was pruned to eliminate relatedness based on identity by descent (pi\_hat $<$ 0.2) \citep{purcell_plink_2007, chang_second-generation_2015}. We curated the genotype data of the EstBB SCZ case/control cohort derived from the GSA array (Illumina’s GSA-MD-24v1, GSA-MD-24v2, ESTchip1\_GSA-MD-24v2, ESTchip2\_GSA-MD-24v3, 2022-09-14 snapshot). Genotype data quality control was performed according to best practices. Specifically, individuals with call rate $<95\%$, who deviated $±$3SD from the samples’ heterozygosity rate mean or showed mismatch between heterozygosity of the X chromosome and sex based on phenotype data were excluded; all AT and GC SNPs, invariable SNPs, SNPs showing potential traces of batch bias, poor cluster separation results and inconsistent allele frequency among any of the EstBB genotyping experiments were removed. The final set contained 290,522 SNPs across 22 autosomal chromosomes after further removing SNPs unique to at least one version of the array, indels and filtering for bi-allelic regions. 

\subsection*{Simulations}
We obtained simulated phenotypes from the same combined SCZ case/control genotype dataset. Instead of simulating genotypes, we simulated phenotypes to keep genome distribution as realistic as possible. More in detail, let $X$ represent the genotype matrix where $X_{ij}$ is the genotype at position $j$ for individual $i$ with possible values -1 (AA), 0 (AB), 1 (BB); $\beta$ the vector of genetic weights sampled from standard normal distribution $N$(0,1); $\epsilon_i$ the noise term sampled from $N$(0, $k \sigma^2$) where $k$ is a scaling factor and $\sigma^2$ is the variance of the total genetic effects. Then the continuous phenotype $Y$ for individual $i$ is defined as:
\[
Y_i = \sum_{j \in D} f^D_j(X_{ij}) + \sum_{j \in R} f^R_j(X_{ij}) + \sum_{(j,k) \in I} f_{jk}(X_{ij}, X_{ik}) + \sum_{(j,k,l) \in I} f_{jkl}(X_{ij}, X_{ik}, X_{il}) + \epsilon_i
\]
where the initial four terms define the genetic effects based on dominant, recessive, two-way (between 2 SNPs) and three-way (between 3 SNPs) interactions; $D$, $R$ and $I$ are disjoint sets of genomic positions for dominant, recessive and interaction effects. For dominant and recessive effect positions, contribution to the phenotype is defined by functions $f_j^D$ and $f_j^R$:

\[
f^D_j(X_{ij}) = \begin{cases} 
\beta_j \cdot X_{ij} & \text{if } X_{ij} \neq \text{-1} \\
0 & \text{otherwise}
\end{cases}
\]

\[
f^R_j(X_{ij}) = \begin{cases} 
\beta_j \cdot X_{ij} & \text{if } X_{ij} = \text{1} \\
0 & \text{otherwise}
\end{cases}
\]

Two-way and three-way interaction effects are defined by:
\[
f_{jk}(X_{ij}, X_{ik}) = \beta_{jk} \cdot X_{ij} \cdot X_{ik}
\]
\[
f_{jkl}(X_{ij}, X_{ik}, X_{il}) = \beta_{jkl} \cdot X_{ij} \cdot X_{ik} \cdot X_{il}
\]

Given a defined threshold $\tau$, the continuous phenotype is then binarized by:

\[
B_i = \begin{cases}
1 & \text{if } Y_i > \tau \\
0 & \text{otherwise}
\end{cases}
\]

In practice, we chose $\tau$ to have approximately the same number of cases (\ensuremath{\sim}1814) as the real SCZ cohort. We simulated a total of 6 scenarios with different noise scaling factors ($k$ = 1, $k$ = 2, $k$ = 3) and two different numbers of causal SNPs (n = 100 and n = 1000) where all four genetic effect types (dominant, recessive, two-way and three-way interactive) had positions with ratios 5:5:4:6, respectively (i.e., for n = 100: 25, 25, 20, 30 and for n = 1000: 250, 250, 200, 300), yielding equal numbers of causal SNPs for dominant/recessive versus interactive effect types. Genotypes were represented by -1, 0, 1 instead of 0, 1, 2 to provide equal importance to either allele in a given position (considering that the neural network weights are initialised with Kaiming uniform distribution, including negative and positive values uniformly). Nevertheless, our preliminary trials with 0, 1, 2 representation provided similar results in terms of model training.

\subsection*{Neural network model}
Models trained for binary trait classification were all feedforward fully connected neural networks with an input layer, two hidden layers and an output layer. The specific architecture consists of an initial dropout layer to randomly mask a proportion of features (p = 0.99) at each batch training; a linear layer with input size = 290,522 and output size = 290; a dropout layer (p = 0.6) followed by ReLU activation; a linear layer with input size = 290 and output size = 29; a dropout layer (p = 0.6) followed by GELU activation; a linear layer with input size = 29 and output size = 1 followed by sigmoid activation function. This equates to total 84,260,139 trainable parameters. Training was performed using cross-entropy loss function and Adam optimizer with learning rate = 1e-5 and weight decay = 1e-3 for regularisation. Initially, we implemented a weighted loss function to address the class imbalance in the dataset; however, this approach was ineffective (i.e., none of the training runs showed a reduction in validation loss). As an alternative, we duplicated the case samples and reduced the control sample size by half through random subsampling to balance the case/control ratio \citep{buda_systematic_2018}. We additionally added random noise $N$(0,0.1) both to class labels (case label = 1, control label = 0) and input genotypes to increase generalizability. All models were trained for 1000 epochs with batch size = 256 using 150 case and 150 control samples as validation. All models were coded with python-3.9 and pythorch-1.11 \citep{paszke_pytorch_2019}.

\subsection*{Interpretability methods}
While there is a plethora of different interpretability methods for deep learning models, we focused on three post hoc and model-agnostic approaches to be able to produce a general framework suitable for most architectures. Two of them, gradients w.r.t input (saliency maps, SM) \citep{simonyan_deep_2014} and integrated gradients (IG) \citep{sundararajan_axiomatic_2017}, are based on gradients and applicable to any differentiable model. SM creation is a commonly used and straightforward approach where gradients w.r.t input are obtained to assess which features in a given input are important for the output probability. For a binary classification neural network model $f$ and input vector $x$, the saliency (with absolute values) for the $j^{th}$ feature is defined as:
\[
SM_j(x) = \left| \frac{\partial f(x)}{\partial x_j} \right|
\]

IG is an improvement over the simpler gradient approach where gradients of the model's output with respect to the input are integrated along a path from a baseline input to the actual input. For a neural network model $f$, input vector $x$, baseline $x’$ and scalar $\alpha \in [0, 1]$, integrated gradient for the $j^{th}$ feature is defined as:
\[
IG_j(x) = (x_j - x'_j) \int_{0}^{1} \frac{\partial f(x' + \alpha (x - x'))}{\partial x_j} \, d\alpha
\]
As the authors of the method emphasise, the choice of baseline vector is important so that a meaningful interpolation can be achieved in the given domain. In our application, we tested various baselines and decided on using a vector of zeros (where the input is a case genotype) which in preliminary analyses, provided better outcomes (i.e., lower FP and higher TP counts) compared to \textbf{(i)} [input = case and control genotypes] - [baseline = vector of average allele dosages for the whole dataset] and \textbf{(ii)} [input = case genotypes] - [baseline = control genotypes] settings. In this regard, our choice of genotype representations with -1, 0, 1 might also have been helpful, since a vector of zeros would correspond to a completely heterozygous genome, without preference over any allele in a given position. We used Captum-0.6.0 python library \citep{kokhlikyan_captum_2020} for calculating SM and IG.

Permutation-based feature importance (PM) is another model-agnostic technique (in this case, the model does not even need to be differentiable) that measures the impact of feature perturbations on model output \citep{breiman_random_2001}. For a neural network model $f$, input vector $x$ and perturbed input vector $x^{(j)}$ (where the $j^{th}$ feature is replaced by a random feature), the permutation feature importance for the $j^{th}$ feature is defined as:
\[
PM_j(x) = \left| f(x) - f(x^{(j)}) \right|
\]
Here, we opted to measure the rate of change in predictions for the feature importance instead of a performance metric such as accuracy, as this approach increases sensitivity. This is because each feature is expected to have only a small contribution without causing substantial alteration in the output prediction.

Using these three methods, we obtained absolute feature importance scores for the case samples (both in real and simulated scenarios) for all trained models, applied L1 normalization across genomic positions and averaged the scores over samples. We call this averaged value mean attribution score (MAS). L1 normalization was performed to transform the score of each feature for a given sample into a measure of contribution, enabling comparisons between samples.

\subsection*{Feature importance to potentially associated loci}
\begin{figure}[H]
\centering
\includegraphics[width=1\linewidth]{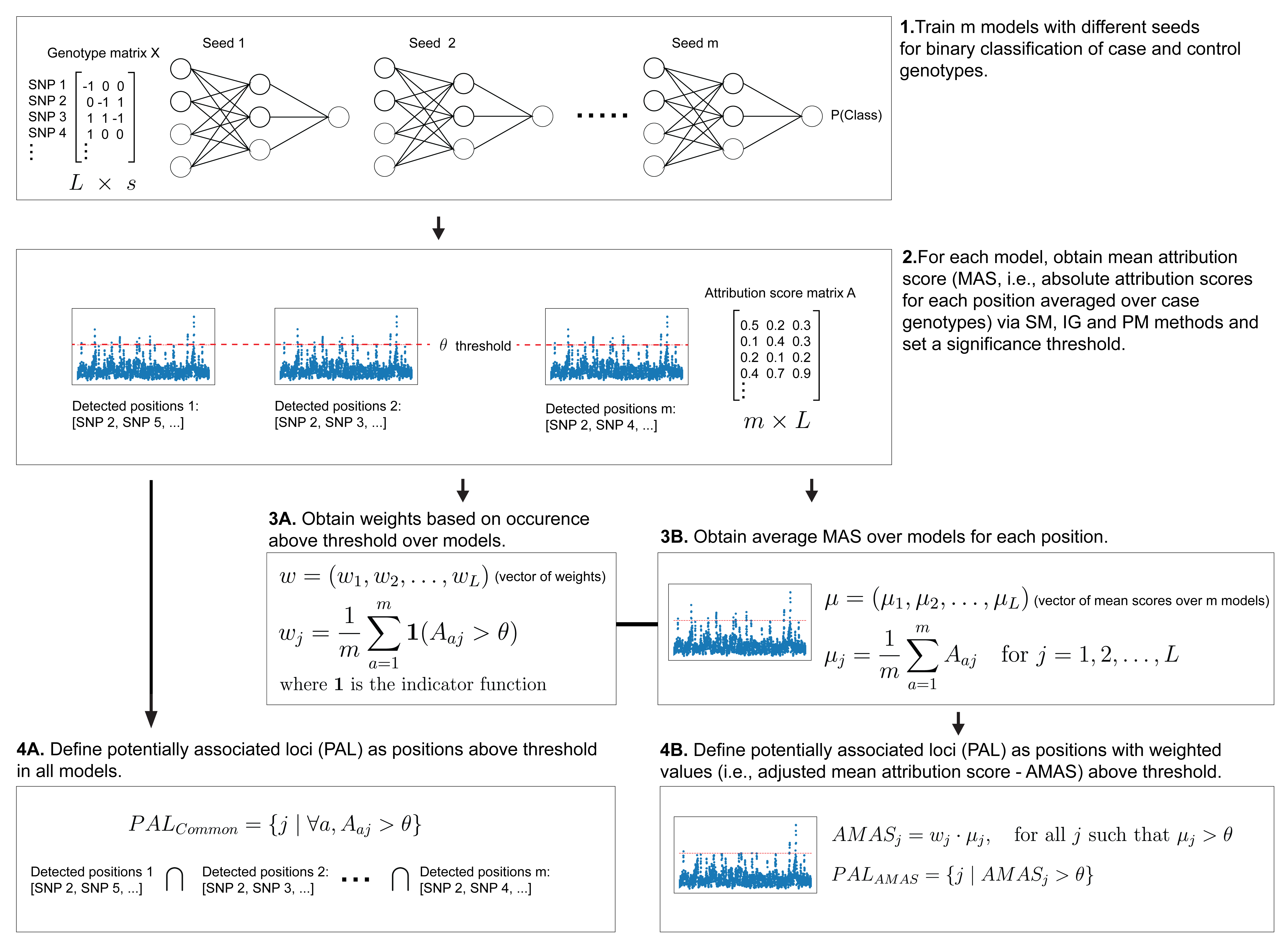}
\caption{Overview of the approach for obtaining potentially associated loci (PAL) from feature attribution scores obtained via SM, IG and PM methods.}
\label{fig:fig1}
\end{figure}

To account for the stochasticity of neural networks and reduce false positive rate, we devised a streamlined method for obtaining potentially associated loci (PAL) from MAS computed from multiple models trained with different seeds. More specifically, let $A$ be the MAS matrix sized $m \times L$ where $m$ is the number of models (i.e., models trained with different seeds) and $L$ is the length of the genotype. We obtain the average MAS vector $\mu$:
\[
\mu = \left(\mu_1, \mu_2, \ldots, \mu_L\right)
\]
where
\[
\mu_j = \frac{1}{m} \sum_{a=1}^m A_{aj} \quad \text{for } j = 1, 2, \ldots, L
\]
Then we set a threshold $\theta$ and obtain weights $w$ based on the number of occurrences above the $\theta$ threshold among $m$ models trained with different seeds:

\[
w_j = \frac{1}{m} \sum_{a=1}^m \mathbf{1}(A_{aj} > \theta), \quad \text{where } \mathbf{1} \text{ is the indicator function}
\]
To obtain PAL, first we define the adjusted mean attribution score (AMAS) as:
\[
AMAS_j = w_j \cdot \mu_j, \quad \text{for all } j \text{ such that } \mu_j > \theta
\]
Then the PAL set contains all positions above the theta threshold based on their AMAS:
\[
PAL_{AMAS} = \left\{ j \mid AMAS_{j} > \theta \right\}
\]
We can also define PAL as positions above the theta threshold in all $m$ models:
\[
PAL_{Common} = \left\{ j \mid \forall a, A_{aj} > \theta \right\}
\]
In practice, we chose $m=10$ for models trained with different seeds to be able to perform a large number of tests with realistic computational resources. Since multiple PAL in high LD were detected by models trained with different seeds, we considered these positions ($r^2>0.5$) across models as common signals to increase detection power. Furthermore, we set the threshold $\theta$ to 99.99 percentile (strict) and 99.95 percentile (relaxed). Although the choice of $\theta$ is essentially arbitrary, the strict threshold yields 30 PAL and the relaxed threshold 145 PAL, which are plausible numbers for maintaining a low false positive rate, given the high polygenicity of schizophrenia \citep{owen_genomic_2023}. An overview of our proposed approach is provided in Fig \ref{fig:fig1}.

\subsection*{Obtaining statistical significance}
The main approach with the $\theta$ threshold weighting is designed for mitigating random effects of stochasticity (i.e., a few ``rogue" models presenting high scoring non-associated loci) but it does not provide a statistical framework for controlling false positives. A common approach for this purpose would be a complete permutation-based method but since this would be computationally infeasible in our case, we instead devised a method to obtain p-values based on the general assumption that higher attribution values are more likely to indicate a true signal. Initially, we trained 10 models with permuted labels (i.e., shuffled binary disease status) and obtained null MAS for each, using the same steps as described previously. Since MAS is calculated as the absolute mean over multiple samples, a reasonable expectation (due to central limit theorem) for the null distribution would be a folded normal distribution with $\mu=0$, which was approximately satisfied in IG null MAS at the tail region of interest (S1 Fig). We estimated the scale parameter $\sigma = 4.2\mathrm{e}{-6}$ from the pooled null MAS and sampled 10 different MAS vectors from the fitted half-normal distribution. To compute AMAS for both simulated and real datasets, we reordered each of the 10 sampled null MAS vectors to match the exact ranking of each of the observed 10 MAS vectors. This procedure emulates dependencies both within (i.e., due to LD and neural network connectivity) and between vectors (i.e., due to shared training data and architecture). We then averaged sampled null MAS vectors and calculated AMAS as previously described. This sampling procedure was repeated 100 times, and p-values were derived by counting the number of sampled AMAS values exceeding the target AMAS. The pseudocode for the method is provided below.

\begin{algorithm}[H]
\caption{Deriving $p$-values for AMAS}
\begin{algorithmic}[1]
\REQUIRE MAS matrix $A$ of size $m \times L$ (with $m$ models and $L$ loci), significance threshold $\theta$, and number of sampled iterations $R$.

\STATE \textbf{Step 1: Compute Observed AMAS}
\STATE Compute the average MAS vector: $\mu_j = \frac{1}{m} \sum_{a=1}^m A_{aj}$ for $j = 1, \dots, L$.
\STATE Compute weights: $w_j = \frac{1}{m} \sum_{a=1}^m \mathbb{I}(A_{aj} > \theta)$, where $\mathbb{I}$ is the indicator function.
\STATE Compute observed AMAS: $\mathrm{AMAS}_j = w_j \cdot \mu_j$ for all $j$ such that $\mu_j > \theta$.

\STATE \textbf{Step 2: Generate Null Distribution}
\FOR{$r = 1$ to $R$}
    \STATE Sample $m$ null MAS vectors from $\text{foldedNormal}(0, \sigma)$, forming the null MAS matrix $A_{\text{null}}^{(r)}$ of size $m \times L$.
    \FOR{$a = 1$ to $m$}
        \STATE Sort the null MAS values for model $a$ to match the rank order of the observed MAS values for the same model:

         \[
        A_{a, :, \text{null}}^{(r)} \leftarrow \text{sort}(A_{a, :, \text{null}}^{(r)},\text{ by order of } A_{a, :})
        \]
        
    \ENDFOR
    \STATE Compute the average null MAS vector: $\mu_j^{(r)} = \frac{1}{m} \sum_{a=1}^m A_{aj,\text{null}}^{(r)}$.
    \STATE Compute null weights: $w_j^{(r)} = \frac{1}{m} \sum_{a=1}^m \mathbb{I}(A_{aj,\text{null}}^{(r)} > \theta)$.
    \STATE Compute null AMAS: $\mathrm{AMAS}_j^{(r)} = w_j^{(r)} \cdot \mu_j^{(r)}$ for all $j$ such that $\mu_j^{(r)} > \theta$.
\ENDFOR

\STATE \textbf{Step 3: Compute $p$-values}
\STATE Combine all null AMAS values across all loci and iterations into a single pooled null distribution.
\FOR{$j = 1$ to $L$}
    \STATE Compute $p$-values as:
    \[
    p_j = \frac{\sum_{r=1}^R \sum_{k=1}^L \mathbb{I}(\mathrm{AMAS}_k^{(r)} > \mathrm{AMAS}_j)}{R \cdot L}
    \]
    
\ENDFOR

\RETURN $p_j$ for all $j$.
\end{algorithmic}
\end{algorithm}

\subsection*{Conventional GWAS}
To compare our method to conventional linear GWAS, we initially performed logistic regression (LR) using all case (n = 1814) and control (n = 7325) samples (both in real and simulated scenarios) using the top 3 principal components (PCs) from principal component analysis (PCA) as covariates. LR without PC covariates did not affect the results (as there is no or limited population structure in the EstBB dataset) (S2 Fig). We used python-3.9 and statsmodels 0.14.1 \citep{seabold_statsmodels_2010} for the LR code. Furthermore, we also performed an association analysis using a state-of-the-art LMM-based tool REGENIE v2.2.4 \citep{mbatchou_computationally_2021} on real data via the exact protocol described by Pujol Gualdo et al. \citep{pujol_gualdo_genome-wide_2023}, except for the filtering steps to retain the same individuals and genomic positions as LR.

Relevant code for the neural network model, interpretability methods and logistic regression with mock simulated data can be found at \url{https://github.com/genodeco/Interpreting-ANNs-for-GWAS}.

\subsection*{Functional analyses}
For PAL detected with the SCZ dataset, we defined genes in the $\pm100$kb region of the detected SNPs as positionally mapped genes. For expression analyses and eQTL mapping, we used the variant-gene association dataset (GTEx\_Analysis\_v8\_eQTL) of the open access Adult Genotype Tissue Expression (GTEx) Project \citep{aguet_genetic_2017}. For all SNPs in PAL, we assessed the significance of  expression levels (significance defined as pval\_nominal $<$ pval\_nominal\_threshold in significant variant-gene pairs dataset from GTEx  *.signif\_variant\_gene\_pairs.txt.gz files) in all brain tissues, and reported eQTL mapped genes for each PAL as genes for which the expression was altered significantly. To assess the functionality of positionally mapped genes in the detected PAL, we used the GENE2FUNC tool in FUMA \citep{watanabe_functional_2017}, which leverages tissue-specific expression levels and phenotype enrichment based on the GWAS catalog \citep{cerezo_nhgri-ebi_2025}. Additionally, to assess the function of each variant, we used the g:Profiler g:SNPense tool \citep{kolberg_gprofilerinteroperable_2023}.

Since we did not perform fine-mapping and multiple SNPs in detected PAL can be due to the same signal, we presented findings only related to our lead SNPs in the literature. More specifically, (i) we extracted all our positionally mapped genes from GWAS Catalog, (ii) identified the SNPs mapped to these genes in the catalog and (iii) presented LD between SNPs reported to be associated with a neuropsychiatric disorder and our lead SNPs. This is not an exhaustive approach and rather conservative, since GWAS Catalog often does not include all significant SNPs in a detected region, and SNPs we detected could potentially be related to other SNPs, especially considering our non-imputed array framework.

\section*{Results}
\subsection*{Simulations}
As post hoc approaches which can be applied to any trained neural network models, we evaluated the performance of the saliency map (SM), integrated gradient (IG) and permutation-based (PM) feature importance methods across six different phenotype simulations, which varied by noise scaling factors ($k$ = 1, $k$ = 2, $k$ = 3) and the number of causal SNPs (100 and 1000). We designated a signal as true positive (TP) if one or more SNPs within a detected PAL block (i.e., blocks formed by clumping detected SNPs less than $100$kb distance) were in close proximity ($\pm100$kb, approximately $\pm20$ SNPs) to a causal position. This approach both accounts for the difficulty of pinpointing the exact causal SNP due to LD between proximal SNPs, and also mitigates bloated TP counts due to multiple signals from the same LD blocks. The results showed that the IG method excels over different simulation settings with the strict $\theta$ threshold in terms of controlling false positives (FPs, i.e., type I errors) compared to other methods, and also performs fairly well in low and moderate noise settings with the relaxed $\theta$ threshold (S3 Fig).
Since IG had better control of FPs and also satisfied the assumption of half-normal null distribution for obtaining p-values (see Materials and Methods, S1 Fig), we compared IG to logistic regression (LR) under the same simulation settings for different significant p-value thresholds, in which 1.7e-07 is the stringent Bonferroni threshold obtained by dividing p-value = 0.05 by the total number of SNPs (Fig \ref{fig:fig2} and S4 Fig). We performed comparisons for different p-value thresholds since multiple testing correction is not straightforward for neural network models (see Discussion). The IG method controlled for FPs better than LR for most simulation parameters and significance thresholds, albeit generally presenting fewer TPs. We further compared the types of causal SNPs (dominant/recessive versus interactive) detected by these two approaches (S5 Fig). Interestingly, mainly positions with interactive genetic effects were detected and there was no apparent difference between the IG and LR method in this aspect. Finally, we visualized the performance of IG on Manhattan plots obtained from LR, demonstrating the capacity of both approaches for detecting multiple causal SNPs (S6 Fig).

\begin{figure}[H]
\centering
\includegraphics[width=1\linewidth]{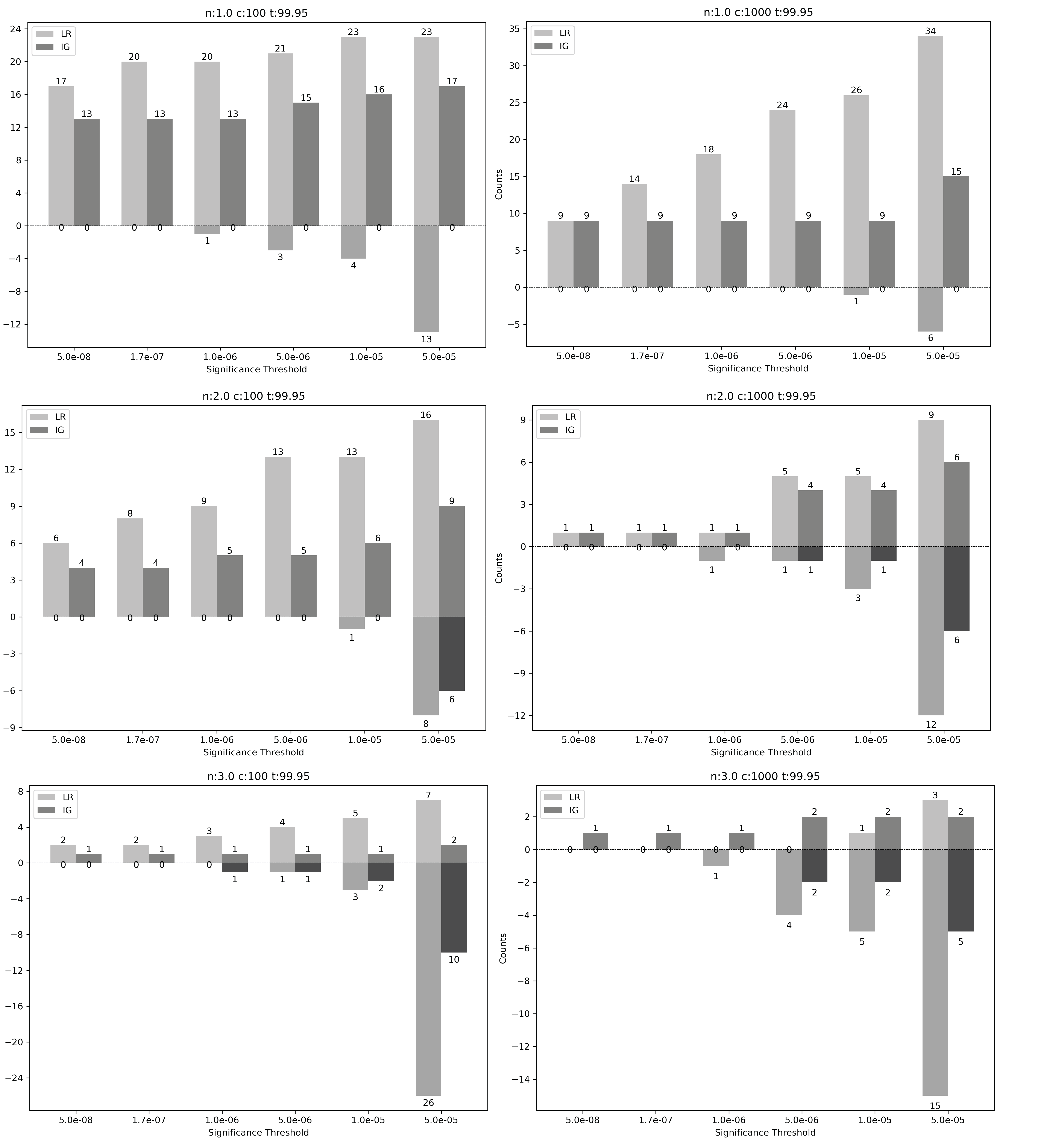}
\caption{Comparison of true positive (TP) and false positive (FP) counts for the integrated gradient (IG) and logistic regression (LR) methods based on various simulation scenarios (n: noise factor, c: number of causal positions, t: relaxed $\theta$ threshold) and different significant p-value thresholds. A signal was determined to be TP if one or more SNP in a detected PAL block (i.e., blocks formed by clumping detected SNPs less than $100$kb distance) was in close proximity ($\pm100$kb, approximately $\pm20$ SNPs) with a causal position. Positive values (above 0 on the y-axis) indicate TP counts whereas negative values (below 0 on the y-axis) indicate FP counts.}
\label{fig:fig2}
\end{figure}

\subsection*{Application to the EstBB SCZ cohort}
After testing the approaches using simulations, we applied the IG method on the EstBB SCZ cohort to detect PAL. We opted not to use SM and PM on real data due to their higher false positive rates under certain simulation scenarios (S3 Fig) and difficulty of obtaining p-values (see Materials and Methods). We detected a single locus with the strict $\theta$ threshold and multiple loci with the relaxed $\theta$ threshold in genomic regions with multiple protein-coding genes (Fig \ref{fig:fig2}, Table \ref{tab:lead_snp_data}). The signal with the highest confidence (p-val $\lessapprox 5 \times 10^{-8}$) on chromosome 11 (PAL f) has been found to be associated with bipolar disorder (a genetically closely related trait with SCZ \citep{romero_exploring_2022}),  with the loci reported in literature including all the genes we report with positional and eQTL mapping \citep{sklar_large-scale_2011}. The lead SNP (rs7944999) in this region is in LD ($r^2 = 0.33$) with SCZ sub-threshold SNP rs1892928 (p-value = 6.8e-07, within \textit{SYT12}) reported in Trubetskoy et al. meta-analysis \citep{trubetskoy_mapping_2022}. Furthermore, it is also in LD ($r^2 = 0.32$) with rs10896135 located in \textit{C11orf80}, a 98 kb open reading frame, and rs7122539 ($r^2 = 0.46$) in \textit{PC}, which are similarly associated with bipolar disorder \citep{sklar_large-scale_2011, stahl_genome-wide_2019}.  Detected PAL b is a gene-dense region found to be associated with mood disturbance (depressive, psychotic and manic symptoms, bipolar II and major depressive disorder), with almost complete overlap with our positionally and eQTL mapped genes \citep{mallard_multivariate_2022}. This region includes rs4955417 ($r^2 = 0.42$ with PAL b lead SNP), which has been associated with depressive symptoms and neuroticism, and was mapped to \textit{IHO1} and \textit{C3orf62} \citep{turley_multi-trait_2018, baselmans_multivariate_2019}), and rs7617480 ($r^2 = 0.33$ with PAL b lead SNP), linked with major depressive disorder and unipolar depression, and mapped to \textit{KLHDC8B} \citep{levey_bi-ancestral_2021,meng_multi-ancestry_2024, als_depression_2023}. Furthermore, \textit{QRICH1} gene within this PAL and reported in the Schizophrenia Exome Sequencing Meta-Analysis (SCHEMA) with p-value$<$0.01, has been linked to developmental disorders and autism spectrum disorder. This region also includes \textit{IMPDH2} and \textit{IP6K2}, both reported in SCHEMA with p-value$<$0.05 \citep{schork_genome-wide_2019, kaplanis_exome-wide_2019, trubetskoy_mapping_2022, chen_shared_2024}. The PAL d region on chromosome 4 contains multiple variants linked to neuropsychiatric disorders. For instance, rs7674220 ($r^2 = 0.28$ with PAL d lead SNP) has been associated with SCZ and mapped to \textit{TET2} \citep{lam_pleiotropic_2019}. Additionally, rs10010325 ($r^2 = 0.48$) (mapped to \textit{TET2}), rs2726528 ($r^2 = 0.48$), rs2713871 ($r^2 = 0.36$) and rs1603705 ($r^2 = 0.20$) (all mapped to \textit{PPA2}), which have been reported to have pleiotropic effects for attention deficit hyperactivity disorder, autism spectrum disorder and intelligence \citep{rao_genetic_2022}. \textit{TET2} gene was also reported in SCHEMA with p-value$<$0.05. An interesting signal we detected is PAL g on chromosome 12, which we positionally mapped to \textit{NTF3}. \textit{NTF3} encodes Neurotrophin-3 protein which has been shown to be linked to SCZ in multiple studies \citep{hattori_novel_2002, arabska_schizophrenia_2018}, but with no reported SNP in GWAS Catalog for SCZ. 

In addition, we assessed the aggregate protein-coding genes we detected in terms of enrichment in phenotypes based on GWAS Catalog (S7 Fig). Lowest enrichment p-value phenotypes were brain morphology and volume, with multiple genes also linked with depressive symptoms, mood instability and intelligence. Among detected variants, rs7647812 (\textit{PRKAR2A}), rs6779394 (\textit{USP19}), rs7944999 (\textit{PC}), and rs10461139, rs10010325, rs6855629, rs2454206 (\textit{TET2}) are targets of nonsense-mediated mRNA decay, while rs4955418 (\textit{CCDC71}) and rs2454206 (\textit{TET2}) are missense variants. The extended information about detected PAL and SNPs is provided in S1 Table.   

\begin{table}[ht]
\centering
\footnotesize
\caption{Detected PAL with lead SNPs and mapped protein coding genes. Genes encompassing lead SNPs are marked with a star.}
\label{tab:lead_snp_data}
\begin{tabular}{lp{2cm}p{2.5cm}p{1.5cm}p{4cm}p{4cm}}
\toprule
PAL & Lead SNP & Chr:Pos & P-val & Positional Mapped (±100 kb) & eQTL Mapped \\ 
\midrule
a & rs4915842 & chr1:59988397 & $1.2 \times 10^{-5}$ & \textit{C1orf87*}, \textit{CYP2J2} & NaN \\ 
b & rs6779394 & chr3:49120338 & $7.8 \times 10^{-6}$ & \textit{C3orf62}, \textit{C3orf84}, \textit{CCDC71},\newline\textit{DALRD3},  \textit{GPX1}, \textit{IHO1},\newline\textit{IMPDH2}, \textit{IP6K2},\newline\textit{KLHDC8B}, \textit{LAMB2}, \newline\textit{NDUFAF3}, \textit{PRKAR2A},\newline \textit{QARS1}, \textit{QRICH1}, \textit{RHOA},\newline\textit{SLC25A20}, \textit{USP19*}, \textit{USP4} & \textit{AMT}, \textit{CCDC71}, \textit{DALRD3}, \newline \textit{GMPPB}, \textit{GPX1}, \newline \textit{KLHDC8B}, \textit{NCKIPSD}, \newline \textit{NICN1}, \textit{P4HTM}, \textit{QRICH1}, \textit{WDR6}\\ 
c & rs12642383 & chr4:28192894 & $3.5 \times 10^{-5}$ & NaN & NaN \\ 
d & rs10461139 & chr4:105184142 & $3.4 \times 10^{-6}$ & \textit{PPA2}, \textit{TET2*} & NaN \\ 
e & rs7454792 & chr6:82072955 & $1.9 \times 10^{-5}$ & \textit{IBTK} & NaN \\ 
f & rs7944999 & chr11:66920392 & $ \lessapprox 5 \times 10^{-8} $ & \textit{C11orf80}, \textit{C11orf86}, \newline \textit{LRFN4}, \textit{PC*}, \textit{RCE1}, \textit{SYT12} & \textit{PC} \\ 
g & rs11063650 & chr12:5403667 & $3.4 \times 10^{-5}$ & \textit{NTF3} & NaN \\ 
\bottomrule
\end{tabular}
\end{table}

\begin{figure}[H]
\centering
\includegraphics[width=1\linewidth]{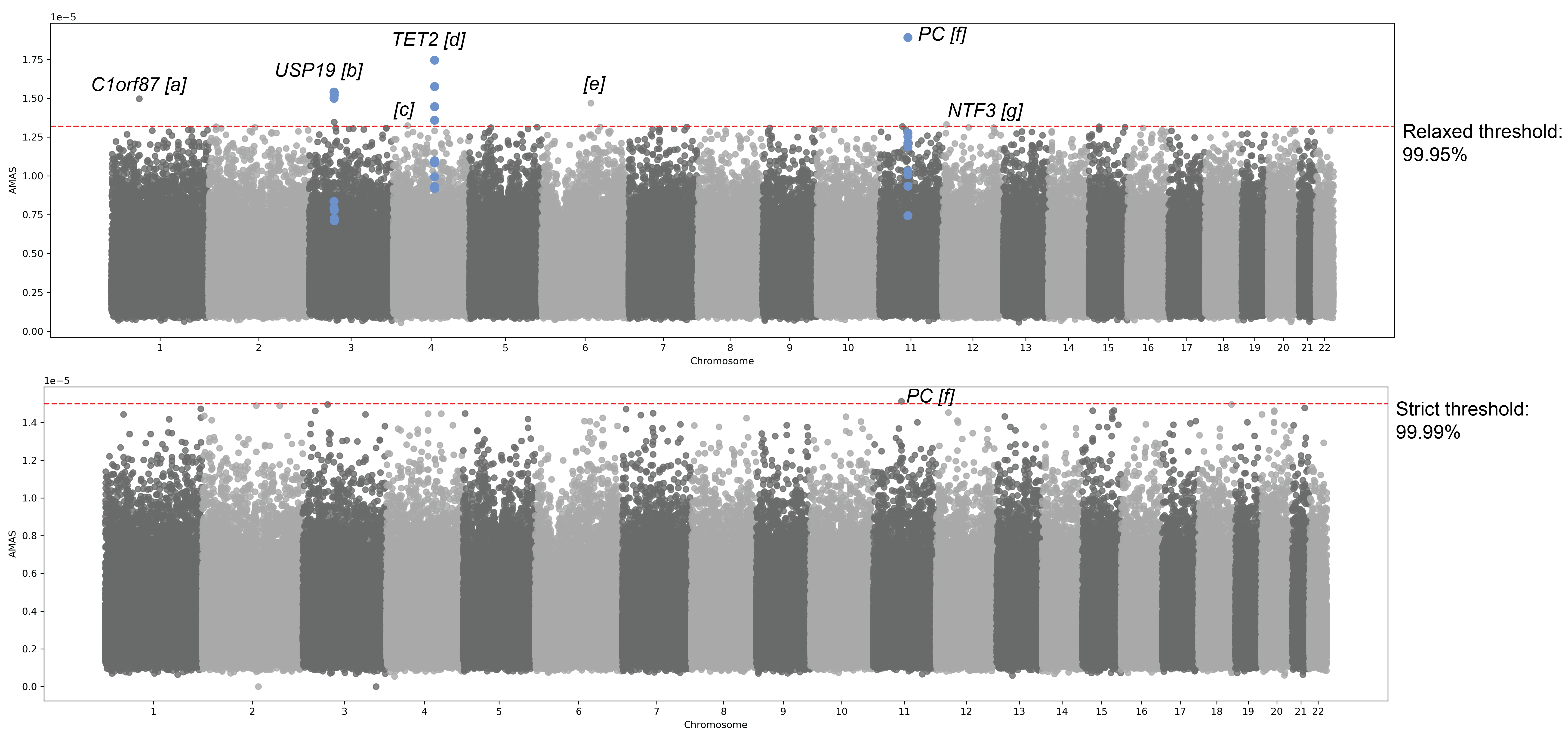}
\caption{PAL detected by the integrated gradients (IG) approach. Red dashed lines indicate $\theta$ thresholds (relaxed and strict) and blue markers indicate PAL above threshold over all trained 10 models (i.e., $PAL_{Common}$). The closest protein coding gene to the lead SNP in $\pm100$kb  region was provided for PAL (a-g).}
\label{fig:fig2}
\end{figure}

We further assessed whether variants in detected PAL were enriched in genes expressed in brain. We initially obtained the empirical cumulative distribution for significant gene-tissue (only brain tissues) expression change in all 290,522 positions. The mean value for all positions was 0.858 (i.e., on average, a SNP had $<1$ gene-tissue combinations with significantly altered expression in brain) with most positions (250,692 SNPs) having no variants significantly affecting gene expression in brain tissues. However, the percentile rank for the detected PAL was 99.4, indicating potential enrichment in brain expressed genes in these detected regions (S8 Fig). As an additional step, we assessed expression specificity for the set of genes within the detected PAL and found significant downregulation in multiple brain tissues (S9 Fig).


Finally, we compared our approach to conventional GWAS using LR and REGENIE association analyses on the real dataset. Although there was high correlation between MAS and -log(p) values (S10 Fig), there was no significant signal detected using LR or REGENIE, and no overlap in signals at lower p-value significance thresholds (S11 Fig). These conventional GWAS results are in line with findings from the UK Biobank, which used a similarly sized case sample set and reported no genome-wide significant hits \citep{legge_genetic_2024}.

\section*{Discussion}

\textbf{Method assessment and simulations.} In this work, we assessed post hoc neural network interpretability methods for detecting trait-associated genomic loci and proposed a general framework for using these methods within a genotype-phenotype context. Simulations demonstrated that these approaches can yield low FP rates with a strict $\theta$ threshold, even in scenarios with relatively weak genetic signals. In addition, the method we proposed for obtaining p-values was robust not only to different scenarios but also to different $\theta$ thresholds for controlling FPs (Fig \ref{fig:fig2} and S4 Fig). An important point to note is that the noise factor of 3 in the simulations corresponds approximately to the SNP-based heritability estimate for SCZ, which is around 0.25 \citep{trubetskoy_mapping_2022}. However, this estimate is based on linear additive models, whereas our simulated traits were based solely on interactive and dominant/recessive effect positions, possibly making detection of associated loci substantially more difficult. Indeed, the neural network model was most successfully trained with the real dataset (i.e., smallest validation loss and least overfitting on average), which in addition increases confidence for the findings from the real dataset (S12 Fig). This connects to a fundamental challenge in method development for detecting causal variants - the absence of ground truth of the underlying model for a trait (i.e., genetic architecture and the interactive landscape of genetic and environmental effects). Without knowing the underlying model, assessing these methods is challenging due to the likely discrepancy between simulated and real data domains. Another challenge for method assessment arises from LD. Even in simulations with ground truth information, evaluation metrics such as TP/FP counts can be substantially biased since methods (including LR) tend to produce highly correlated signals. We aimed to mitigate this with our definition of TP (i.e., if one or more SNP in a detected block is in $\pm100$kb region of a causal position), but methods with different capacities for detecting signals in LD regions might still be difficult to compare reliably. Future research could benefit from evaluation metrics better tailored for genomic data.

\textbf{Uncertainty quantification.} An important obstacle to the broader adoption of the proposed approaches for GWAS lies in the difficulty of establishing reliable statistical significance measures. While our approach for obtaining p-values effectively controlled FPs in difficult simulation settings, a stable multiple testing correction remains to be a challenge due to complex dependencies stemming from the neural network architecture and gradient calculations. Furthermore, we used a pooled approach to obtain p-values based on a global null distribution for all features, essentially testing for relative extremity. This is fundamentally different from conventional GWAS where each SNP is tested against its own null distribution. The global null approach allowed us to obtain meaningful p-values (i.e., can be corrected considering thousands of SNPs) via a feasible number of sampling iterations but it also inherently limits the detection of small but significant effects. Another important limitation is that it might not be possible to estimate the null distribution for all types of feature importance scores (S1 Fig), which is especially relevant if a complete permutation approach with multiple training runs is not computationally feasible as in our application. As a conceptually different solution for controlling FPs, knockoff variables have been adopted in previous works \citep{candes_panning_2018, kassani_deep_2022, badre_explainable_2023}. Briefly, these are implanted “fake” positions which mimic the distribution of real input features and enable the estimation of FP rates due to their independence from the output variable. However, sampling knockoff variables is not straightforward because deviations from the distribution of real input features would bias the false positive estimates. Furthermore, implementing a sufficient number of knockoff variables would potentially deteriorate the realness of data and significantly increase the computational burden due to the fully connected architecture of our model. Future parameter-light architectures (such as convolutions) or prior feature selection could be more suitable to explore this approach.

\textbf{Linear versus non-linear models.} Another important subject related to neural networks is model training. In our approach, we employed both heavy dropout regularisation (including unusual dropout masking before the input layer) and ensemble-like multiple seed training to take advantage of stochasticity and limit FPs consequently. Interestingly, even with this heavy regularisation, we were able to train models successfully (i.e., positive correlation between training and validation loss), even though the predictive power of these models was limited. Receiver operating characteristic (ROC) curves for phenotype prediction in the real dataset were comparable between neural network and logistic regression models, with logistic regression being marginally better (S13 Fig). Initially, this might suggest that the real dataset primarily exhibits linear effects. Yet, as a counter-evidence to the “mainly linear effects” scheme, logistic regression or REGENIE could not detect any significant positions when applied to the real cohort, and there was no correspondence between the top detected positions of linear and neural network approaches (S11 Fig), which might suggest the existence of nonlinearities in the real dataset. These aspects of effect types and how heavy regularisation alters the capacity of neural networks for detecting nonlinearities are interesting subjects for future research. Considering the relatively small dataset for a case/cohort type analysis in this work, increasing the training data and experimenting with more simulation scenarios (such as including more nonlinearities, gene-environment interactions or covariate effects) could be helpful to distinguish the differences between linear and nonlinear methodologies.

\textbf{Interactive and dominant/recessive effects.} One intriguing finding was that all approaches mainly detected positions with interactive effects. This may be due to dominant/recessive positions being more difficult to detect since the SNP effect completely disappears for a proportion of genotypes (see Materials and Methods). Additionally, the outcome space of functions defining complex trait genomics is considerably smaller than that of continuous variables, such as gene expression measurements, due to the ternary nature of genotype variables. Research into the topology of this space could provide valuable insights about the limitations of methods developed for in silico detection of causal loci, and provide better roadmaps and improvement strategies.

\textbf{Application to real datasets.} In SCZ cohort application, we presented extensive evidence based on literature and enrichment analyses to demonstrate that the detected loci are related to neuropsychiatric disorders and brain function. Future research should focus on replicating these results for different populations, biobanks and phenotypes. Furthermore, to mitigate confounding factors, we distribution matched case and control datasets based on age, sex and BMI, and pruned the data to eliminate cryptic relatedness. Since EstBB dataset do not exhibit high population structure (e.g., LR findings did not change when PCs are used as covariates, S2 Fig), we did not include PCs as input features for the neural network model. Future models could integrate potential confounding variables to improve generalizability over different datasets and reduce data preprocessing steps.

\textbf{Final remarks.} Despite the state-of-the-art deep learning algorithms and the ever-increasing amount of data in biobanks, several challenges remain in the wide-spread adoption of neural networks for GWAS. These include stochasticity stemming from initial random states, lack of established uncertainty quantification and the absence of ground truth for the underlying genetic and environmental landscape of complex traits. These challenges are fundamentally connected to intriguing questions, such as the contribution of nonlinear effects (gene-gene or gene-environment interactions or other polynomial disease models) or omnigenic versus polygenic models for complex disease development \citep{boyle_expanded_2017}. In the near future, assumption-free interpretable neural networks may become essential tools for exploring these questions and could ultimately replace linear models as the standard approach for GWAS and the generation of interpretable polygenic risk scores.

\section*{Acknowledgements}
BY, MA, MNG, LM were supported by the European Union’s Horizon 2020 Research and Innovation Programme under Grant agreements No 964874 Realment and 847776 CoMorMent. Thanks to the Agence Nationale de la Recherche (ANR-20-CE45-0010-01) for funding. Thanks to the High Performance Computing Center (HPC) of the University of Tartu for providing computational resources. Thanks to the Estonian Biobank Research Team (Andres Metspalu, Tõnu Esko, Reedik Mägi, Mari Nelis, Georgi Hudjashov) for the genotype and phenotype datasets used in this work. Thanks to Liis Karo-Astover for administrative support and proofreading. Data used for the gene-variant expression analyses described in this manuscript were obtained from the Genotype-Tissue Expression (GTEx) Portal on 04/04/2024.

\section*{Contributions}
Conceptualisation: BY. Data curation: BY, MA. Methodology: BY. Investigation: BY, MA, FJ. Formal analysis: BY, MA, MNG. Visualisation: BY. Supervision: BY, FJ, LM. Original draft: BY. Review \& editing: BY, MA, MNG, FJ, LM. Funding acquisition: LM, FJ, BY.

\section*{Ethics statement}
The activities of the EstBB are regulated by the Human Genes Research Act, which was adopted in 2000 specifically for the operations of the EstBB. Individual level data analysis was carried out under ethical approval 1.1-12/624 (issued 24.03.2020), 1.1-12/2618 (issued 04.08.2022) and 1.1-12/3454 (issued 20.10.2022) from the Estonian Committee on Bioethics and Human Research (Estonian Ministry of Social Affairs).

\section*{Supplementary Table Legends}

\textbf{S1 Table.} Detected PAL including all SNPs with mapped genes.


\bibliography{library}

\begin{thebibliography}{10}

\bibitem{abdellaoui_15_2023}
A.~Abdellaoui, L.~Yengo, K.~J.~H. Verweij, and P.~M. Visscher, ``15 years of {GWAS} discovery: {Realizing} the promise,'' {\em The American Journal of Human Genetics}, vol.~110, pp.~179--194, Feb. 2023.

\bibitem{jiang_generalized_2021}
L.~Jiang, Z.~Zheng, H.~Fang, and J.~Yang, ``A generalized linear mixed model association tool for biobank-scale data,'' {\em Nature Genetics}, vol.~53, pp.~1616--1621, Nov. 2021.
\newblock Publisher: Nature Publishing Group.

\bibitem{mbatchou_computationally_2021}
J.~Mbatchou, L.~Barnard, J.~Backman, A.~Marcketta, J.~A. Kosmicki, A.~Ziyatdinov, C.~Benner, C.~O’Dushlaine, M.~Barber, B.~Boutkov, L.~Habegger, M.~Ferreira, A.~Baras, J.~Reid, G.~Abecasis, E.~Maxwell, and J.~Marchini, ``Computationally efficient whole-genome regression for quantitative and binary traits,'' {\em Nature Genetics}, vol.~53, pp.~1097--1103, July 2021.
\newblock Publisher: Nature Publishing Group.

\bibitem{lu_universal_2020}
Y.~Lu and J.~Lu, ``A universal approximation theorem of deep neural networks for expressing probability distributions,'' in {\em Proceedings of the 34th {International} {Conference} on {Neural} {Information} {Processing} {Systems}}, {NIPS} '20, (Red Hook, NY, USA), pp.~3094--3105, Curran Associates Inc., Dec. 2020.

\bibitem{alharbi_review_2022}
W.~S. Alharbi and M.~Rashid, ``A review of deep learning applications in human genomics using next-generation sequencing data,'' {\em Human Genomics}, vol.~16, p.~26, July 2022.

\bibitem{zhou_deep_2023}
X.~Zhou, Y.~Chen, F.~C.~F. Ip, Y.~Jiang, H.~Cao, G.~Lv, H.~Zhong, J.~Chen, T.~Ye, Y.~Chen, Y.~Zhang, S.~Ma, R.~M.~N. Lo, E.~P.~S. Tong, V.~C.~T. Mok, T.~C.~Y. Kwok, Q.~Guo, K.~Y. Mok, M.~Shoai, J.~Hardy, L.~Chen, A.~K.~Y. Fu, and N.~Y. Ip, ``Deep learning-based polygenic risk analysis for {Alzheimer}’s disease prediction,'' {\em Communications Medicine}, vol.~3, pp.~1--20, Apr. 2023.
\newblock Publisher: Nature Publishing Group.

\bibitem{van_hilten_gennet_2021}
A.~van Hilten, S.~A. Kushner, M.~Kayser, M.~A. Ikram, H.~H.~H. Adams, C.~C.~W. Klaver, W.~J. Niessen, and G.~V. Roshchupkin, ``{GenNet} framework: interpretable deep learning for predicting phenotypes from genetic data,'' {\em Communications Biology}, vol.~4, pp.~1--9, Sept. 2021.
\newblock Number: 1 Publisher: Nature Publishing Group.

\bibitem{xu_interpretable_2022}
J.~Xu, C.~Mao, Y.~Hou, Y.~Luo, J.~L. Binder, Y.~Zhou, L.~M. Bekris, J.~Shin, M.~Hu, F.~Wang, C.~Eng, T.~I. Oprea, M.~E. Flanagan, A.~A. Pieper, J.~Cummings, J.~B. Leverenz, and F.~Cheng, ``Interpretable deep learning translation of {GWAS} and multi-omics findings to identify pathobiology and drug repurposing in {Alzheimer}’s disease,'' {\em Cell Reports}, vol.~41, p.~111717, Nov. 2022.

\bibitem{liu_explainable_2022}
L.~Liu, Q.~Meng, C.~Weng, Q.~Lu, T.~Wang, and Y.~Wen, ``Explainable deep transfer learning model for disease risk prediction using high-dimensional genomic data,'' {\em PLOS Computational Biology}, vol.~18, p.~e1010328, July 2022.
\newblock Publisher: Public Library of Science.

\bibitem{badre_explainable_2023}
A.~Badré and C.~Pan, ``Explainable multi-task learning improves the parallel estimation of polygenic risk scores for many diseases through shared genetic basis,'' {\em PLOS Computational Biology}, vol.~19, p.~e1011211, July 2023.
\newblock Publisher: Public Library of Science.

\bibitem{liu_phenotype_2019}
Y.~Liu, D.~Wang, F.~He, J.~Wang, T.~Joshi, and D.~Xu, ``Phenotype {Prediction} and {Genome}-{Wide} {Association} {Study} {Using} {Deep} {Convolutional} {Neural} {Network} of {Soybean},'' {\em Frontiers in Genetics}, vol.~10, 2019.

\bibitem{mieth_deepcombi_2021}
B.~Mieth, A.~Rozier, J.~A. Rodriguez, M.~M.~C. Höhne, N.~Görnitz, and K.-R. Müller, ``{DeepCOMBI}: explainable artificial intelligence for the analysis and discovery in genome-wide association studies,'' {\em NAR Genomics and Bioinformatics}, vol.~3, p.~lqab065, Sept. 2021.

\bibitem{kassani_deep_2022}
P.~H. Kassani, F.~Lu, Y.~Le~Guen, M.~E. Belloy, and Z.~He, ``Deep neural networks with controlled variable selection for the identification of putative causal genetic variants,'' {\em Nature Machine Intelligence}, vol.~4, pp.~761--771, Sept. 2022.
\newblock Publisher: Nature Publishing Group.

\bibitem{sharma_deep_2020}
D.~Sharma, A.~Durand, M.-A. Legault, L.-P.~L. Perreault, A.~Lemaçon, M.-P. Dubé, and J.~Pineau, ``Deep interpretability for {GWAS}.''

\bibitem{leitsalu_cohort_2015}
L.~Leitsalu, T.~Haller, T.~Esko, M.-L. Tammesoo, H.~Alavere, H.~Snieder, M.~Perola, P.~C. Ng, R.~Mägi, L.~Milani, K.~Fischer, and A.~Metspalu, ``Cohort {Profile}: {Estonian} {Biobank} of the {Estonian} {Genome} {Center}, {University} of {Tartu},'' {\em International Journal of Epidemiology}, vol.~44, pp.~1137--1147, Aug. 2015.

\bibitem{milani_biobanking_2024}
L.~Milani, M.~Alver, S.~Laur, {\em et~al.}, ``From biobanking to personalized medicine: the journey of the estonian biobank.''

\bibitem{alver_genetic_2024}
M.~Alver, S.~Kasela, L.~Haring, L.~B. Luitva, K.~Fischer, M.~Möls, and L.~Milani, ``Genetic predisposition and antipsychotic treatment effect on metabolic syndrome in schizophrenia: a ten-year follow-up study using the estonian biobank,'' vol.~41.
\newblock Publisher: Elsevier.

\bibitem{ho_matchit_2011}
D.~Ho, K.~Imai, G.~King, and E.~A. Stuart, ``{MatchIt}: {Nonparametric} {Preprocessing} for {Parametric} {Causal} {Inference},'' {\em Journal of Statistical Software}, vol.~42, pp.~1--28, June 2011.

\bibitem{purcell_plink_2007}
S.~Purcell, B.~Neale, K.~Todd-Brown, L.~Thomas, M.~A.~R. Ferreira, D.~Bender, J.~Maller, P.~Sklar, P.~I.~W. de~Bakker, M.~J. Daly, and P.~C. Sham, ``{PLINK}: A tool set for whole-genome association and population-based linkage analyses,'' vol.~81, no.~3, pp.~559--575.

\bibitem{chang_second-generation_2015}
C.~C. Chang, C.~C. Chow, L.~C. Tellier, S.~Vattikuti, S.~M. Purcell, and J.~J. Lee, ``Second-generation {PLINK}: rising to the challenge of larger and richer datasets,'' vol.~4, no.~1, pp.~s13742--015--0047--8.

\bibitem{buda_systematic_2018}
M.~Buda, A.~Maki, and M.~A. Mazurowski, ``A systematic study of the class imbalance problem in convolutional neural networks,'' {\em Neural Networks}, vol.~106, pp.~249--259, Oct. 2018.

\bibitem{paszke_pytorch_2019}
A.~Paszke, S.~Gross, F.~Massa, A.~Lerer, J.~Bradbury, G.~Chanan, T.~Killeen, Z.~Lin, N.~Gimelshein, L.~Antiga, A.~Desmaison, A.~Köpf, E.~Yang, Z.~DeVito, M.~Raison, A.~Tejani, S.~Chilamkurthy, B.~Steiner, L.~Fang, J.~Bai, and S.~Chintala, ``{PyTorch}: {An} {Imperative} {Style}, {High}-{Performance} {Deep} {Learning} {Library},'' Dec. 2019.
\newblock arXiv:1912.01703 [cs, stat].

\bibitem{simonyan_deep_2014}
K.~Simonyan, A.~Vedaldi, and A.~Zisserman, ``Deep {Inside} {Convolutional} {Networks}: {Visualising} {Image} {Classification} {Models} and {Saliency} {Maps},'' Apr. 2014.
\newblock arXiv:1312.6034 [cs].

\bibitem{sundararajan_axiomatic_2017}
M.~Sundararajan, A.~Taly, and Q.~Yan, ``Axiomatic attribution for deep networks,'' in {\em Proceedings of the 34th {International} {Conference} on {Machine} {Learning} - {Volume} 70}, {ICML}'17, (Sydney, NSW, Australia), pp.~3319--3328, JMLR.org, Aug. 2017.

\bibitem{kokhlikyan_captum_2020}
N.~Kokhlikyan, V.~Miglani, M.~Martin, E.~Wang, B.~Alsallakh, J.~Reynolds, A.~Melnikov, N.~Kliushkina, C.~Araya, S.~Yan, and O.~Reblitz-Richardson, ``Captum: {A} unified and generic model interpretability library for {PyTorch},'' 2020.
\newblock \_eprint: 2009.07896.

\bibitem{breiman_random_2001}
L.~Breiman, ``Random {Forests},'' {\em Machine Learning}, vol.~45, pp.~5--32, Oct. 2001.

\bibitem{owen_genomic_2023}
M.~J. Owen, ``Genomic insights into schizophrenia,'' {\em Royal Society Open Science}, vol.~10, p.~230125, Feb. 2023.
\newblock Publisher: Royal Society.

\bibitem{seabold_statsmodels_2010}
S.~Seabold and J.~Perktold, ``statsmodels: {Econometric} and statistical modeling with python,'' in {\em 9th {Python} in {Science} {Conference}}, 2010.

\bibitem{pujol_gualdo_genome-wide_2023}
N.~Pujol~Gualdo, {Estonian Biobank Research Team}, R.~Mägi, and T.~Laisk, ``Genome-wide association study meta-analysis supports association between {MUC1} and ectopic pregnancy,'' {\em Human Reproduction}, vol.~38, pp.~2516--2525, Dec. 2023.

\bibitem{aguet_genetic_2017}
F.~Aguet, A.~A. Brown, S.~E. Castel, {\em et~al.}, ``Genetic effects on gene expression across human tissues,'' {\em Nature}, vol.~550, pp.~204--213, Oct. 2017.
\newblock Publisher: Nature Publishing Group.

\bibitem{watanabe_functional_2017}
K.~Watanabe, E.~Taskesen, A.~van Bochoven, and D.~Posthuma, ``Functional mapping and annotation of genetic associations with {FUMA},'' {\em Nature Communications}, vol.~8, p.~1826, Nov. 2017.
\newblock Publisher: Nature Publishing Group.

\bibitem{cerezo_nhgri-ebi_2025}
M.~Cerezo, E.~Sollis, Y.~Ji, E.~Lewis, A.~Abid, K.~Bircan, P.~Hall, J.~Hayhurst, S.~John, A.~Mosaku, S.~Ramachandran, A.~Foreman, A.~Ibrahim, J.~{McLaughlin}, Z.~Pendlington, R.~Stefancsik, S.~A. Lambert, A.~{McMahon}, J.~Morales, T.~Keane, M.~Inouye, H.~Parkinson, and L.~W. Harris, ``The {NHGRI}-{EBI} {GWAS} catalog: standards for reusability, sustainability and diversity,'' vol.~53, pp.~D998--D1005.

\bibitem{kolberg_gprofilerinteroperable_2023}
L.~Kolberg, U.~Raudvere, I.~Kuzmin, P.~Adler, J.~Vilo, and H.~Peterson, ``g:profiler—interoperable web service for functional enrichment analysis and gene identifier mapping (2023 update),'' vol.~51, pp.~W207--W212.

\bibitem{romero_exploring_2022}
C.~Romero, J.~Werme, P.~R. Jansen, J.~Gelernter, M.~B. Stein, D.~Levey, R.~Polimanti, C.~de~Leeuw, D.~Posthuma, M.~Nagel, and S.~van~der Sluis, ``Exploring the genetic overlap between twelve psychiatric disorders,'' vol.~54, no.~12, pp.~1795--1802.
\newblock Publisher: Nature Publishing Group.

\bibitem{sklar_large-scale_2011}
P.~Sklar, S.~Ripke, L.~J. Scott, {\em et~al.}, ``Large-scale genome-wide association analysis of bipolar disorder identifies a new susceptibility locus near {ODZ}4,'' vol.~43, no.~10, pp.~977--983.
\newblock Publisher: Nature Publishing Group.

\bibitem{trubetskoy_mapping_2022}
V.~Trubetskoy, A.~F. Pardiñas, T.~Qi, {\em et~al.}, ``Mapping genomic loci implicates genes and synaptic biology in schizophrenia,'' {\em Nature}, vol.~604, pp.~502--508, Apr. 2022.
\newblock Publisher: Nature Publishing Group.

\bibitem{stahl_genome-wide_2019}
E.~A. Stahl, G.~Breen, A.~J. Forstner, {\em et~al.}, ``Genome-wide association study identifies 30 loci associated with bipolar disorder,'' {\em Nature Genetics}, vol.~51, pp.~793--803, May 2019.
\newblock Publisher: Nature Publishing Group.

\bibitem{mallard_multivariate_2022}
T.~T. Mallard, R.~Karlsson~Linnér, A.~D. Grotzinger, S.~Sanchez-Roige, J.~Seidlitz, A.~Okbay, R.~de~Vlaming, S.~F.~W. Meddens, A.~A. Palmer, L.~K. Davis, E.~M. Tucker-Drob, K.~S. Kendler, M.~C. Keller, P.~D. Koellinger, and K.~P. Harden, ``Multivariate {GWAS} of psychiatric disorders and their cardinal symptoms reveal two dimensions of cross-cutting genetic liabilities,'' vol.~2, no.~6, p.~100140.

\bibitem{turley_multi-trait_2018}
P.~Turley, R.~K. Walters, O.~Maghzian, A.~Okbay, J.~J. Lee, M.~A. Fontana, T.~A. Nguyen-Viet, R.~Wedow, M.~Zacher, N.~A. Furlotte, P.~Magnusson, S.~Oskarsson, M.~Johannesson, P.~M. Visscher, D.~Laibson, D.~Cesarini, B.~M. Neale, and D.~J. Benjamin, ``Multi-trait analysis of genome-wide association summary statistics using {MTAG},'' {\em Nature Genetics}, vol.~50, pp.~229--237, Feb. 2018.
\newblock Publisher: Nature Publishing Group.

\bibitem{baselmans_multivariate_2019}
B.~M.~L. Baselmans, R.~Jansen, H.~F. Ip, J.~van Dongen, A.~Abdellaoui, M.~P. van~de Weijer, Y.~Bao, M.~Smart, M.~Kumari, G.~Willemsen, J.-J. Hottenga, D.~I. Boomsma, E.~J.~C. de~Geus, M.~G. Nivard, and M.~Bartels, ``Multivariate genome-wide analyses of the well-being spectrum,'' {\em Nature Genetics}, vol.~51, pp.~445--451, Mar. 2019.
\newblock Publisher: Nature Publishing Group.

\bibitem{levey_bi-ancestral_2021}
D.~F. Levey, M.~B. Stein, F.~R. Wendt, G.~A. Pathak, H.~Zhou, M.~Aslan, R.~Quaden, K.~M. Harrington, Y.~Z. Nuñez, C.~Overstreet, K.~Radhakrishnan, G.~Sanacora, A.~M. McIntosh, J.~Shi, S.~S. Shringarpure, J.~Concato, R.~Polimanti, and J.~Gelernter, ``Bi-ancestral depression {GWAS} in the {Million} {Veteran} {Program} and meta-analysis in {\textgreater}1.2 million individuals highlight new therapeutic directions,'' {\em Nature Neuroscience}, vol.~24, pp.~954--963, July 2021.
\newblock Publisher: Nature Publishing Group.

\bibitem{meng_multi-ancestry_2024}
X.~Meng, G.~Navoly, O.~Giannakopoulou, {\em et~al.}, ``Multi-ancestry genome-wide association study of major depression aids locus discovery, fine mapping, gene prioritization and causal inference,'' {\em Nature Genetics}, vol.~56, pp.~222--233, Feb. 2024.
\newblock Publisher: Nature Publishing Group.

\bibitem{als_depression_2023}
T.~D. Als, M.~I. Kurki, J.~Grove, G.~Voloudakis, K.~Therrien, E.~Tasanko, T.~T. Nielsen, J.~Naamanka, K.~Veerapen, D.~F. Levey, J.~Bendl, J.~Bybjerg-Grauholm, B.~Zeng, D.~Demontis, A.~Rosengren, G.~Athanasiadis, M.~Bækved-Hansen, P.~Qvist, G.~Bragi~Walters, T.~Thorgeirsson, H.~Stefánsson, K.~L. Musliner, V.~M. Rajagopal, L.~Farajzadeh, J.~Thirstrup, B.~J. Vilhjálmsson, J.~J. McGrath, M.~Mattheisen, S.~Meier, E.~Agerbo, K.~Stefánsson, M.~Nordentoft, T.~Werge, D.~M. Hougaard, P.~B. Mortensen, M.~B. Stein, J.~Gelernter, I.~Hovatta, P.~Roussos, M.~J. Daly, O.~Mors, A.~Palotie, and A.~D. Børglum, ``Depression pathophysiology, risk prediction of recurrence and comorbid psychiatric disorders using genome-wide analyses,'' {\em Nature Medicine}, vol.~29, pp.~1832--1844, July 2023.
\newblock Publisher: Nature Publishing Group.

\bibitem{schork_genome-wide_2019}
A.~J. Schork, H.~Won, V.~Appadurai, R.~Nudel, M.~Gandal, O.~Delaneau, M.~Revsbech~Christiansen, D.~M. Hougaard, M.~Bækved-Hansen, J.~Bybjerg-Grauholm, M.~Giørtz~Pedersen, E.~Agerbo, C.~Bøcker~Pedersen, B.~M. Neale, M.~J. Daly, N.~R. Wray, M.~Nordentoft, O.~Mors, A.~D. Børglum, P.~Bo~Mortensen, A.~Buil, W.~K. Thompson, D.~H. Geschwind, and T.~Werge, ``A genome-wide association study of shared risk across psychiatric disorders implicates gene regulation during fetal neurodevelopment,'' {\em Nature Neuroscience}, vol.~22, pp.~353--361, Mar. 2019.
\newblock Publisher: Nature Publishing Group.

\bibitem{kaplanis_exome-wide_2019}
J.~Kaplanis, N.~Akawi, G.~Gallone, J.~F. McRae, E.~Prigmore, C.~F. Wright, D.~R. Fitzpatrick, H.~V. Firth, J.~C. Barrett, M.~E. Hurles, and o.~b. o. t. D. D.~D. Study, ``Exome-wide assessment of the functional impact and pathogenicity of multinucleotide mutations,'' {\em Genome Research}, vol.~29, pp.~1047--1056, July 2019.
\newblock Company: Cold Spring Harbor Laboratory Press Distributor: Cold Spring Harbor Laboratory Press Institution: Cold Spring Harbor Laboratory Press Label: Cold Spring Harbor Laboratory Press Publisher: Cold Spring Harbor Lab.

\bibitem{chen_shared_2024}
Y.~Chen, W.~Li, L.~Lv, and W.~Yue, ``Shared {Genetic} {Determinants} of {Schizophrenia} and {Autism} {Spectrum} {Disorder} {Implicate} {Opposite} {Risk} {Patterns}: {A} {Genome}-{Wide} {Analysis} of {Common} {Variants},'' {\em Schizophrenia Bulletin}, p.~sbae044, Apr. 2024.

\bibitem{lam_pleiotropic_2019}
M.~Lam, W.~D. Hill, J.~W. Trampush, J.~Yu, {\em et~al.}, ``Pleiotropic {Meta}-{Analysis} of {Cognition}, {Education}, and {Schizophrenia} {Differentiates} {Roles} of {Early} {Neurodevelopmental} and {Adult} {Synaptic} {Pathways},'' {\em The American Journal of Human Genetics}, vol.~105, pp.~334--350, Aug. 2019.

\bibitem{rao_genetic_2022}
S.~Rao, A.~Baranova, Y.~Yao, J.~Wang, and F.~Zhang, ``Genetic {Relationships} between {Attention}-{Deficit}/{Hyperactivity} {Disorder}, {Autism} {Spectrum} {Disorder}, and {Intelligence},'' {\em Neuropsychobiology}, vol.~81, pp.~484--496, June 2022.

\bibitem{hattori_novel_2002}
M.~Hattori, H.~Kunugi, A.~Akahane, H.~Tanaka, S.~Ishida, T.~Hirose, R.~Morita, K.~Yamakawa, and S.~Nanko, ``Novel polymorphisms in the promoter region of the neurotrophin-3 gene and their associations with schizophrenia,'' vol.~114, no.~3, pp.~304--309.
\newblock \_eprint: https://onlinelibrary.wiley.com/doi/pdf/10.1002/ajmg.10248.

\bibitem{arabska_schizophrenia_2018}
J.~Arabska, A.~Łucka, D.~Strzelecki, and A.~Wysokiński, ``In schizophrenia serum level of neurotrophin-3 ({NT}-3) is increased only if depressive symptoms are present,'' vol.~684, pp.~152--155.

\bibitem{legge_genetic_2024}
S.~E. Legge, A.~F. Pardiñas, G.~Woolway, E.~Rees, A.~G. Cardno, V.~Escott-Price, P.~Holmans, G.~Kirov, M.~J. Owen, M.~C. O’Donovan, and J.~T.~R. Walters, ``Genetic and phenotypic features of schizophrenia in the {UK} biobank,'' vol.~81, no.~7, pp.~681--690.

\bibitem{candes_panning_2018}
E.~Candès, Y.~Fan, L.~Janson, and J.~Lv, ``Panning for {Gold}: ‘{Model}-{X}’ {Knockoffs} for {High} {Dimensional} {Controlled} {Variable} {Selection},'' {\em Journal of the Royal Statistical Society Series B: Statistical Methodology}, vol.~80, pp.~551--577, June 2018.

\bibitem{boyle_expanded_2017}
E.~A. Boyle, Y.~I. Li, and J.~K. Pritchard, ``An {Expanded} {View} of {Complex} {Traits}: {From} {Polygenic} to {Omnigenic},'' {\em Cell}, vol.~169, pp.~1177--1186, June 2017.

\end{thebibliography}

\bibliographystyle{ieeetr}


\end{document}



\section*{Supplementary Figures}

\setcounter{figure}{0}
\renewcommand{\figurename}{S\arabic{figure}}
\renewcommand{\thefigure}{Fig}

\begin{figure*}[!ht]
\centering
\includegraphics[width=0.7\linewidth]{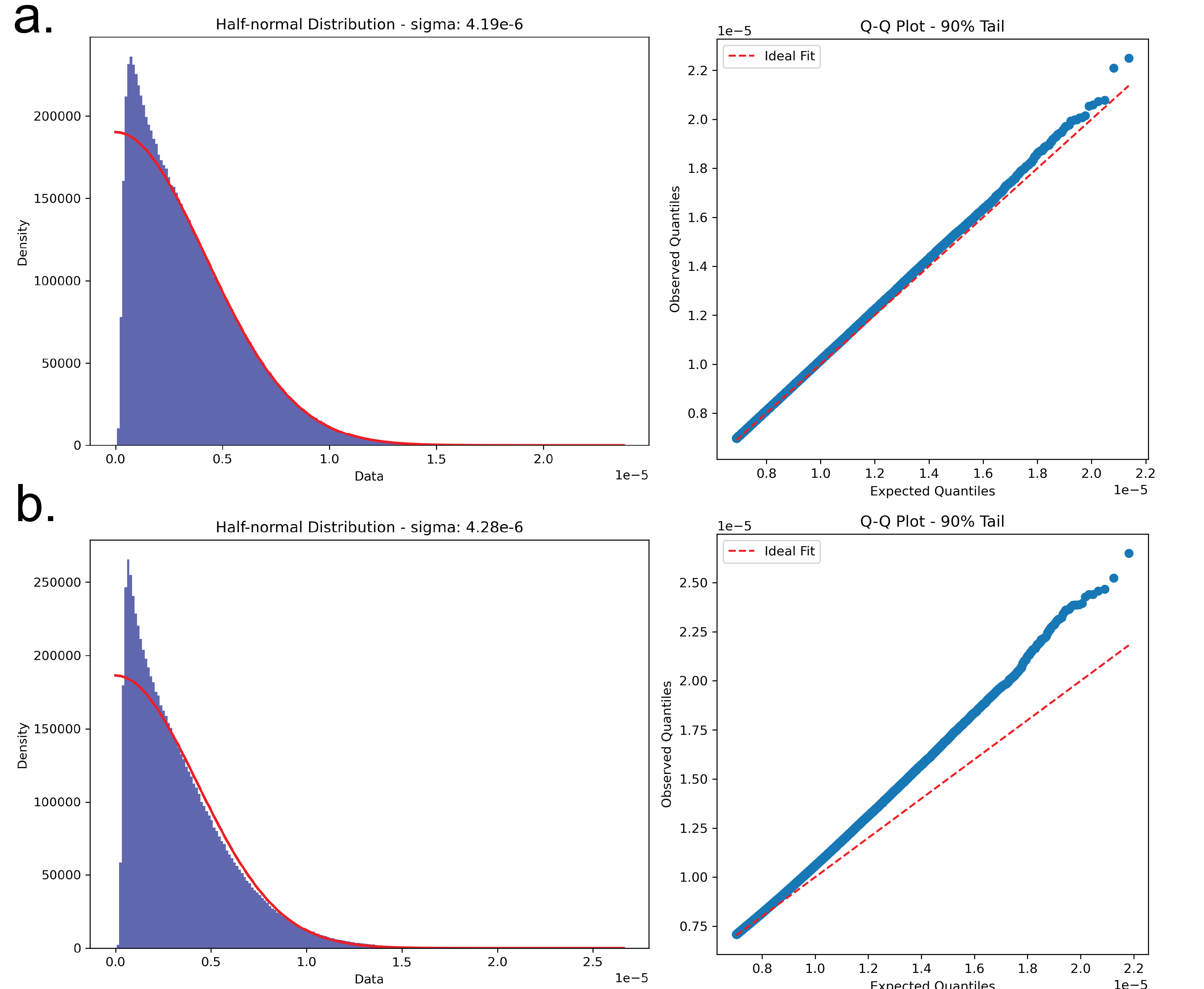}
\caption{Assessment of half-normal distribution fit to \textbf{a)} integrated gradient (IG) and \textbf{b)} saliency map (SM) null mean attribution score (MAS) distribution obtained via 10 model training with permuted labels. Red lines indicate the theoretical fit. Right column is the Q-Q plot of expected versus observed values for the 90th percentile tail.
}
\label{suppfig:supfig1}
\end{figure*}

\begin{figure*}[!h]
\centering
\includegraphics[width=0.5\linewidth]{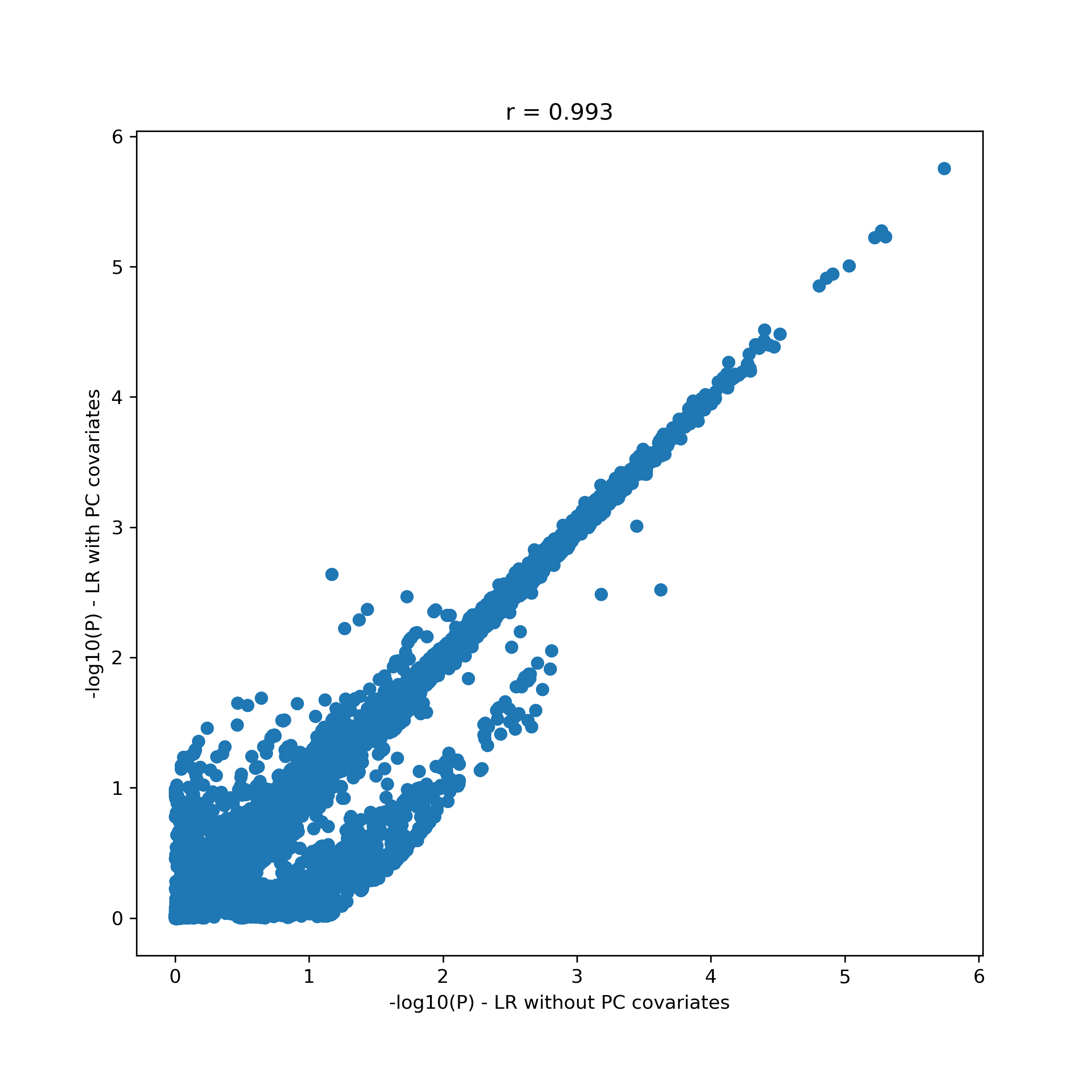}
\caption{Comparison of logistic regression (LR) p-values without (x-axis) and with (y-axis) PC covariates.}
\label{suppfig:supfig1}
\end{figure*}

\begin{figure*}[!ht]
\centering
\includegraphics[width=0.57\linewidth]{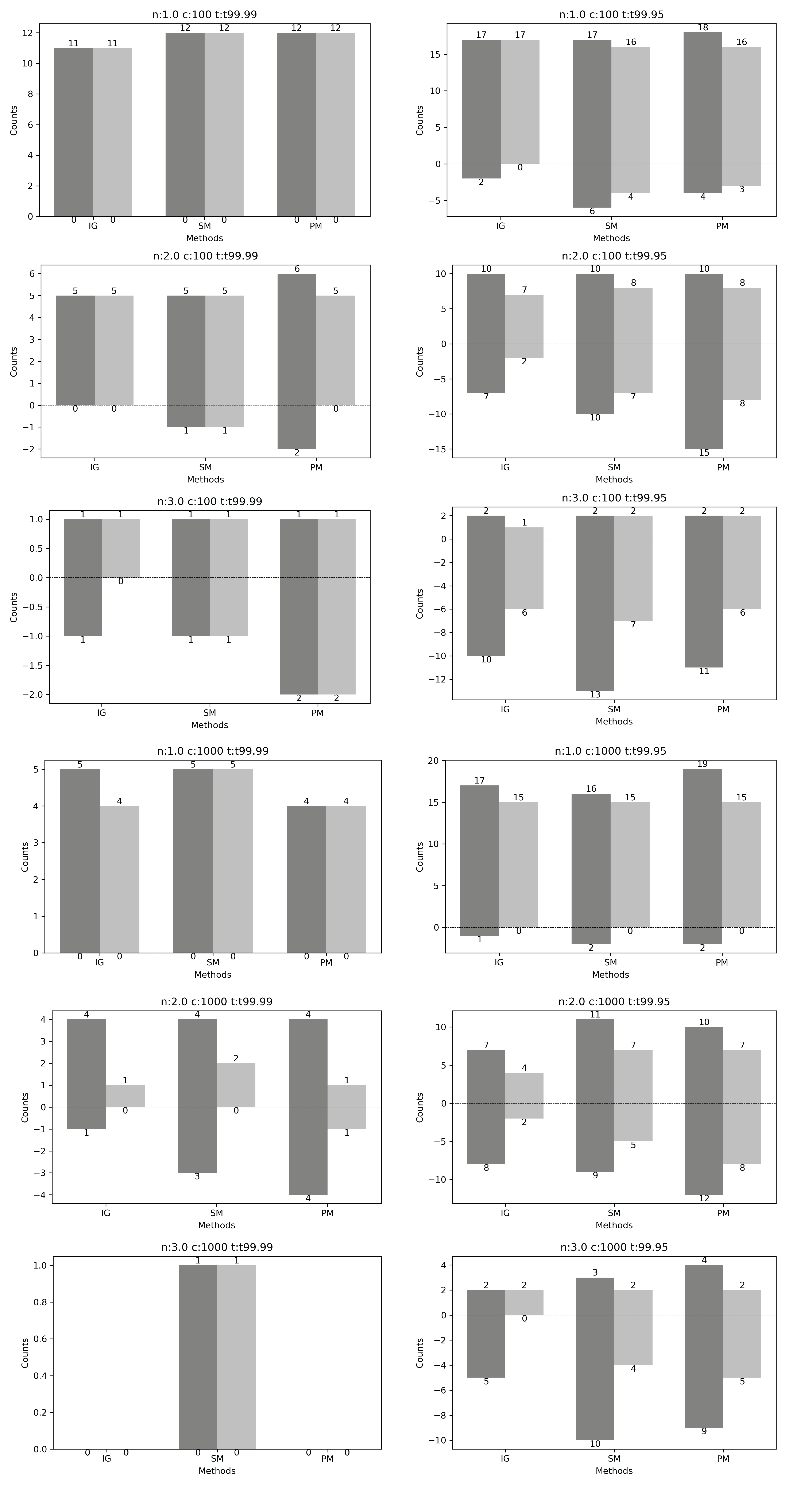}
\caption{Comparison of true positive (TP) and false positive (FP) counts for different methods based on various simulation scenarios and thresholds (n: noise factor, c: number of causal positions, t: $\theta$ threshold). A signal was determined to be TP if one or more SNP in a detected PAL block (i.e., blocks formed by clumping detected SNPs less than $100$kb distance) is in close proximity ($\pm100$kb, approximately $\pm20$ SNPs) with a causal position. Positive values (above 0 on the y-axis) indicate TP counts whereas negative values (below 0 on the y-axis) indicate FP counts. Methods defined in x-axis are integrated gradients (IG), saliency map (SM) and permutation-based (PM) approaches. First bars (dark grey) show TP/FP counts for $PAL_{AMAS}$ whereas second bars (light grey) show TP/FP counts for $PAL_{Common}$.
}
\label{suppfig:supfig1}
\end{figure*}

\begin{figure*}
\centering
\includegraphics[width=1\linewidth]{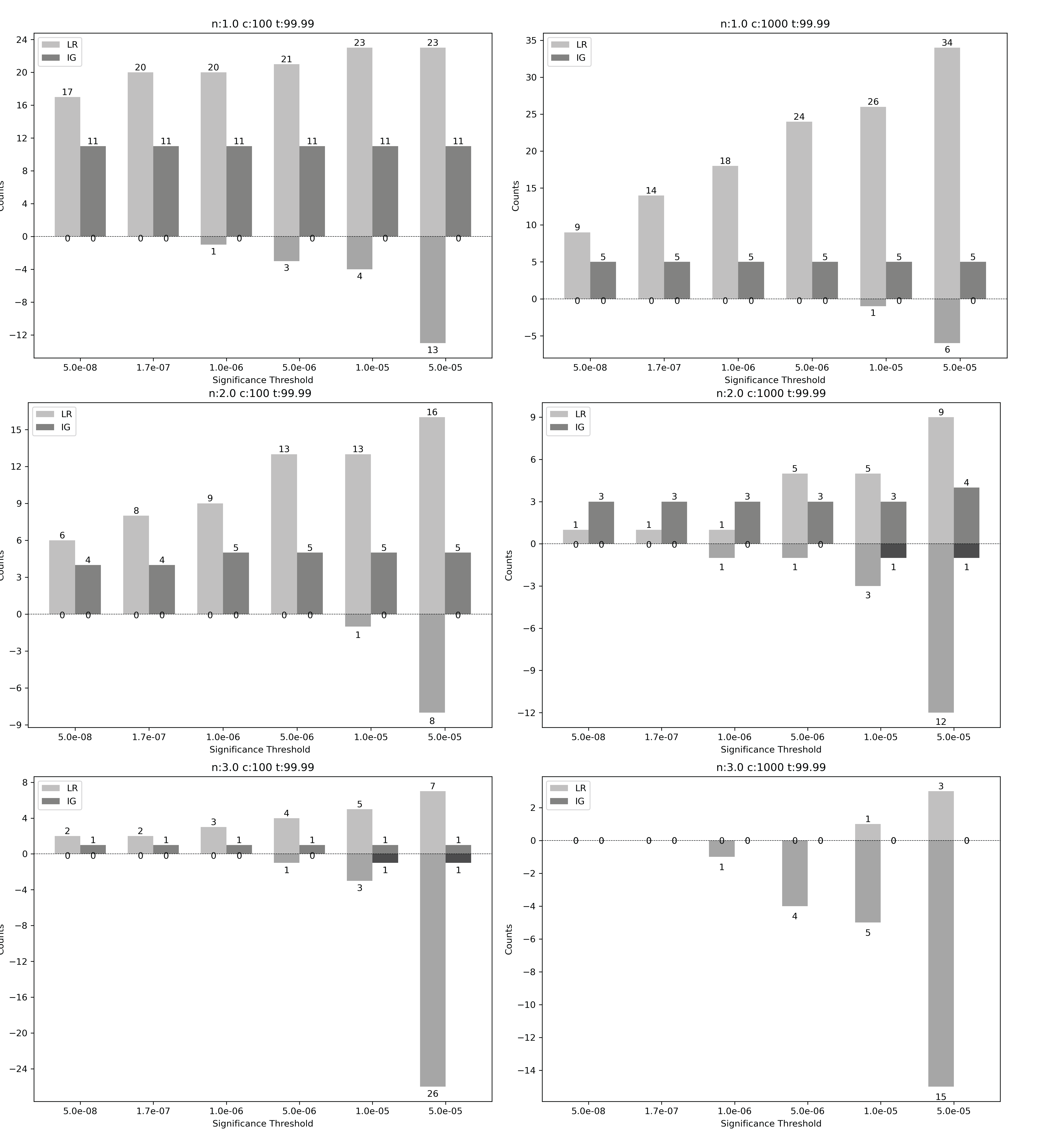}
\caption{Comparison of true positive (TP) and false positive (FP) counts for integrated gradient (IG) and logistic regression (LR) methods based on various simulation scenarios (n: noise factor, c: number of causal positions, t: strict $\theta$ threshold) and different significant p-value thresholds. A signal was determined to be TP if one or more SNP in a detected PAL block (i.e., blocks formed by clumping detected SNPs less than $100$kb distance) is in close proximity ($\pm100$kb, approximately $\pm20$ SNPs) with a causal position. Positive values (above 0 on the y-axis) indicate TP counts whereas negative values (below 0 on the y-axis) indicate FP counts.}
\label{suppfig:supfig3}
\end{figure*}

\begin{figure*}
\centering
\includegraphics[width=1\linewidth]{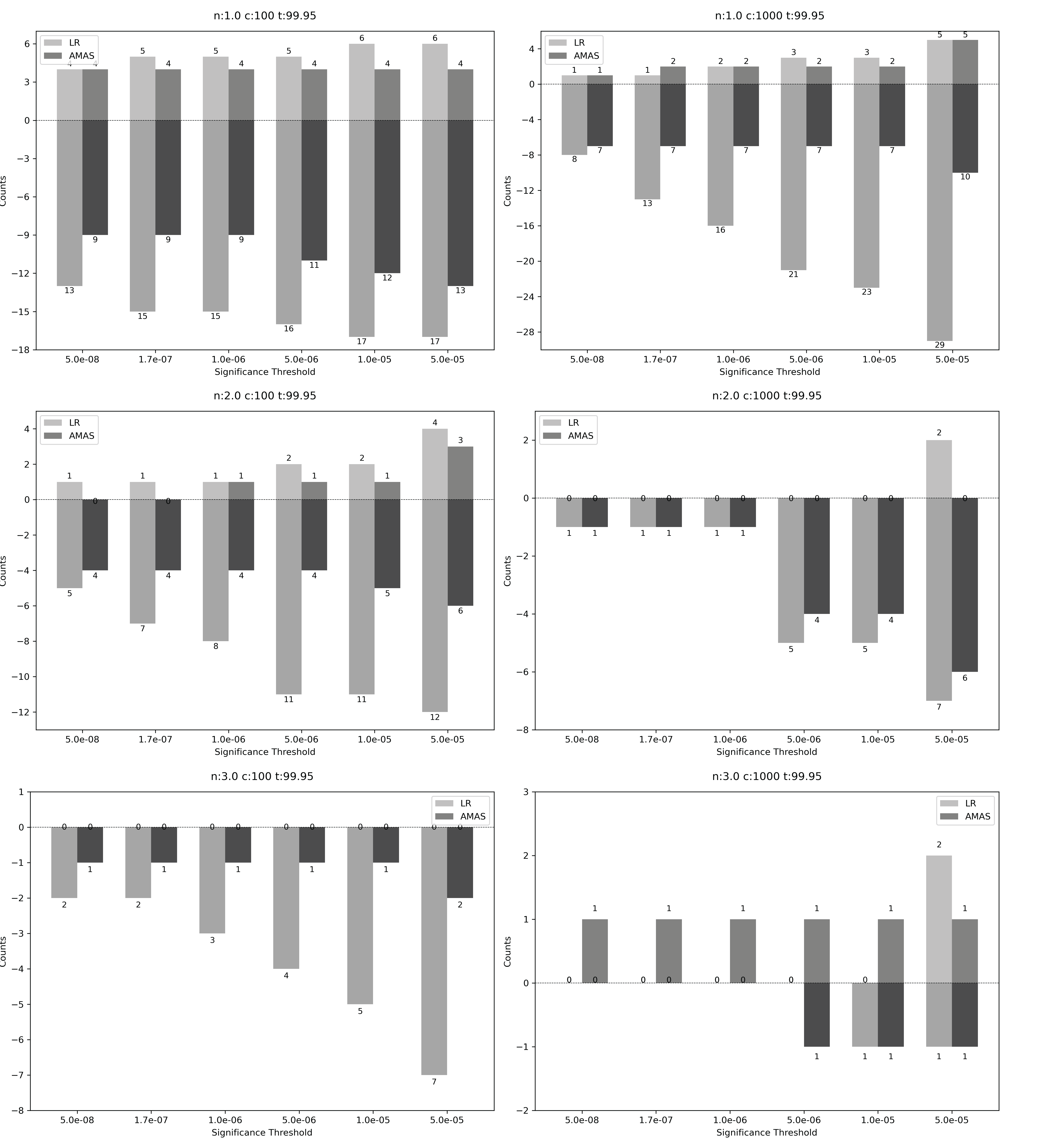}
\caption{Comparison of dominant/recessive (positive values) and interactive effect (negative values) position counts in correctly detected (true positive) positions for integrated gradient (IG) and logistic regression (LR) methods based on various simulation scenarios (n: noise factor, c: number of causal positions, t: relaxed $\theta$ threshold) and different significant p-value thresholds.}
\label{suppfig:supfig3}
\end{figure*}

\begin{figure*}[!ht]
\centering
\includegraphics[width=1\linewidth]{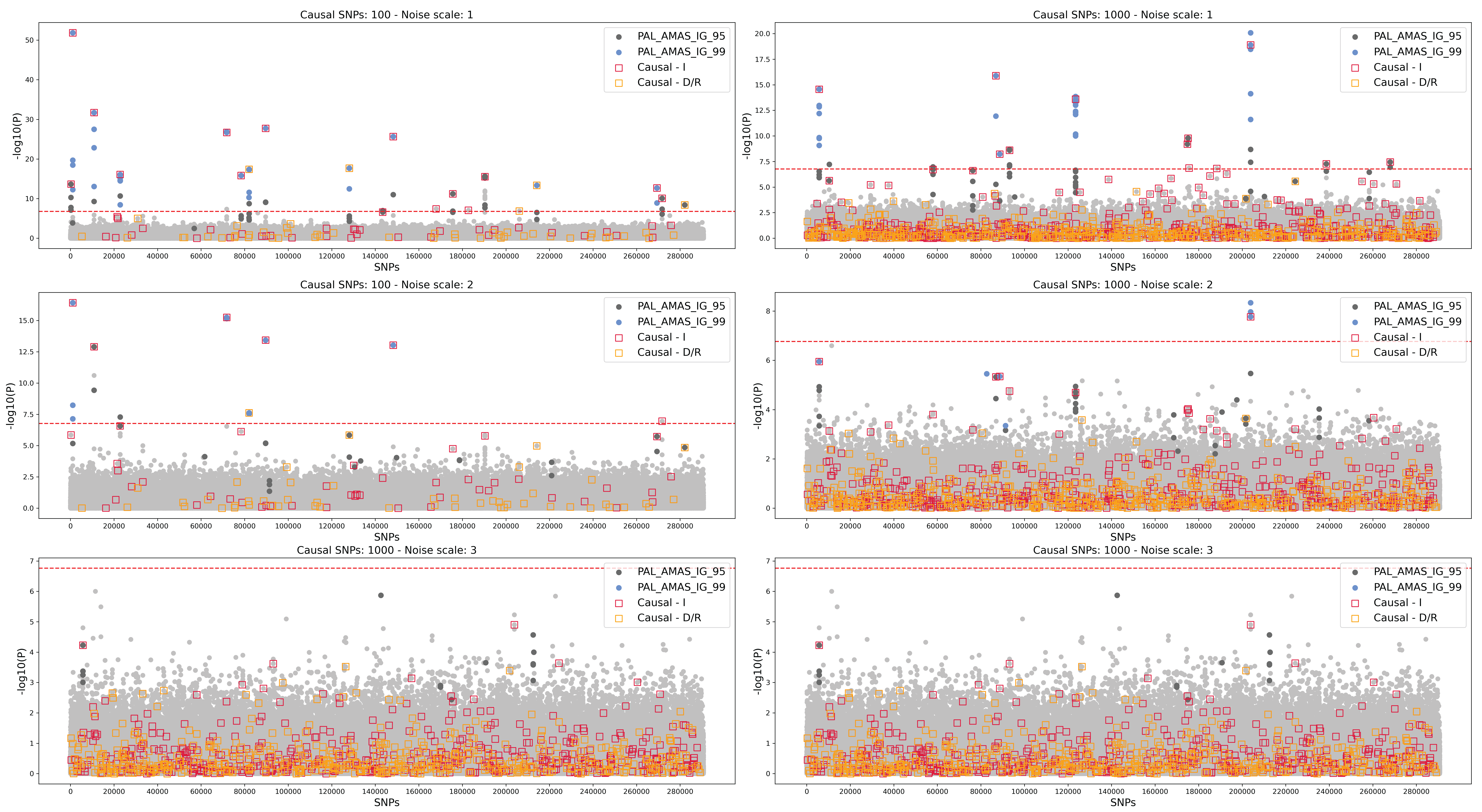}
\caption{Manhattan plots from LR with annotated $PAL_{AMAS}$ detected via IG approach. x-axis denotes SNP positions over the whole genome, and y-axis denotes -log(P) values of the LR analysis. Each figure is a different simulation scenario with varying amounts of noise and causal SNPs, as described in the subcaptions. Dashed red lines denote the significance threshold for LR (with Bonferroni corrected p-value = 0.05 divided by the number of tests). Red squares denote causal SNPs with interactive effects (Causal - I), and orange squares denote causal SNPs with dominant/recessive effects (Causal - D/R). Blue and black dots represent $PAL_{AMAS}$ detected with strict (99.99 percentile) and relaxed (99.95 percentile) $\theta$ thresholds, respectively.}
\label{suppfig:supfig2}
\end{figure*}

\begin{figure*}
\centering
\includegraphics[width=1\linewidth]{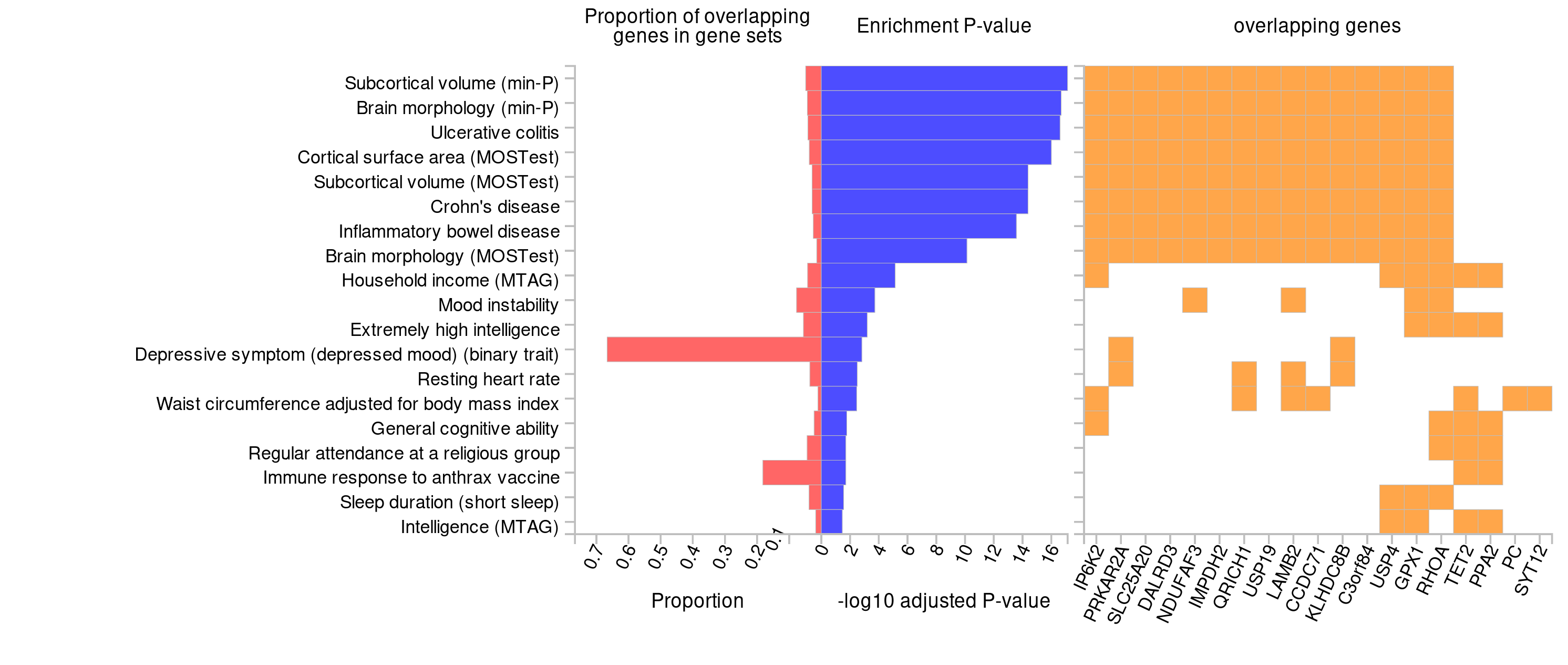}
\caption{FUMA functional analysis of detected genes based on enrichment in GWAS-catalogue gene sets.}
\label{suppfig:supfig4}
\end{figure*}

\begin{figure*}
\centering
\includegraphics[width=1\linewidth]{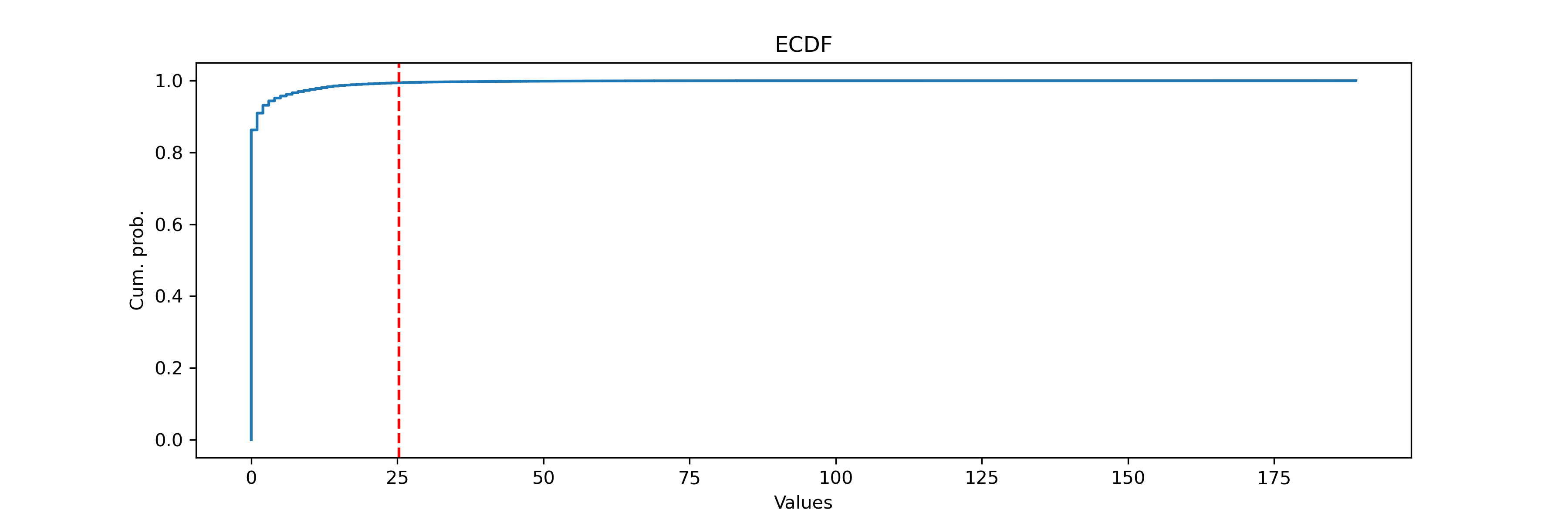}
\caption{Empirical cumulative distribution function (ECDF) plot of brain tissue expression for all 290,522 SNPs. Values (x-axis) correspond to the number of gene-tissue (consisting of all types of brain tissues) combinations with significant expression change. Percentile rank for the average value over detected PAL set is 99.4 (354 significant gene-tissue expression for 14 detected SNPs).}
\label{suppfig:supfig3}
\end{figure*}

\begin{figure*}
\centering
\includegraphics[width=1\linewidth]{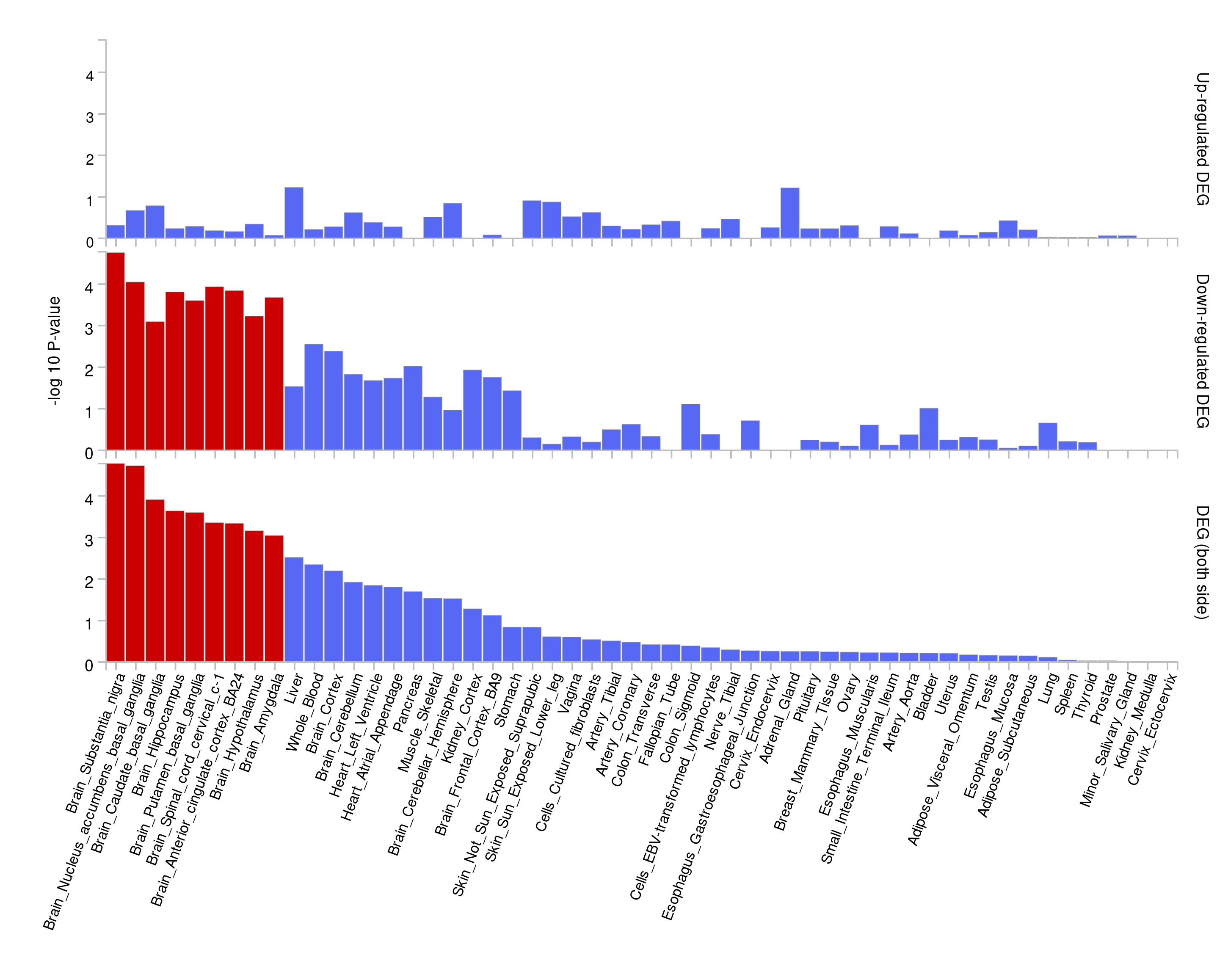}
\caption{FUMA gene expression analysis of detected genes. Differentially expressed gene (DEG) sets with significant enrichments based on Bonferroni correction are coloured in red.}
\label{suppfig:supfig4}
\end{figure*}

\begin{figure*}
\centering
\includegraphics[width=0.6\linewidth]{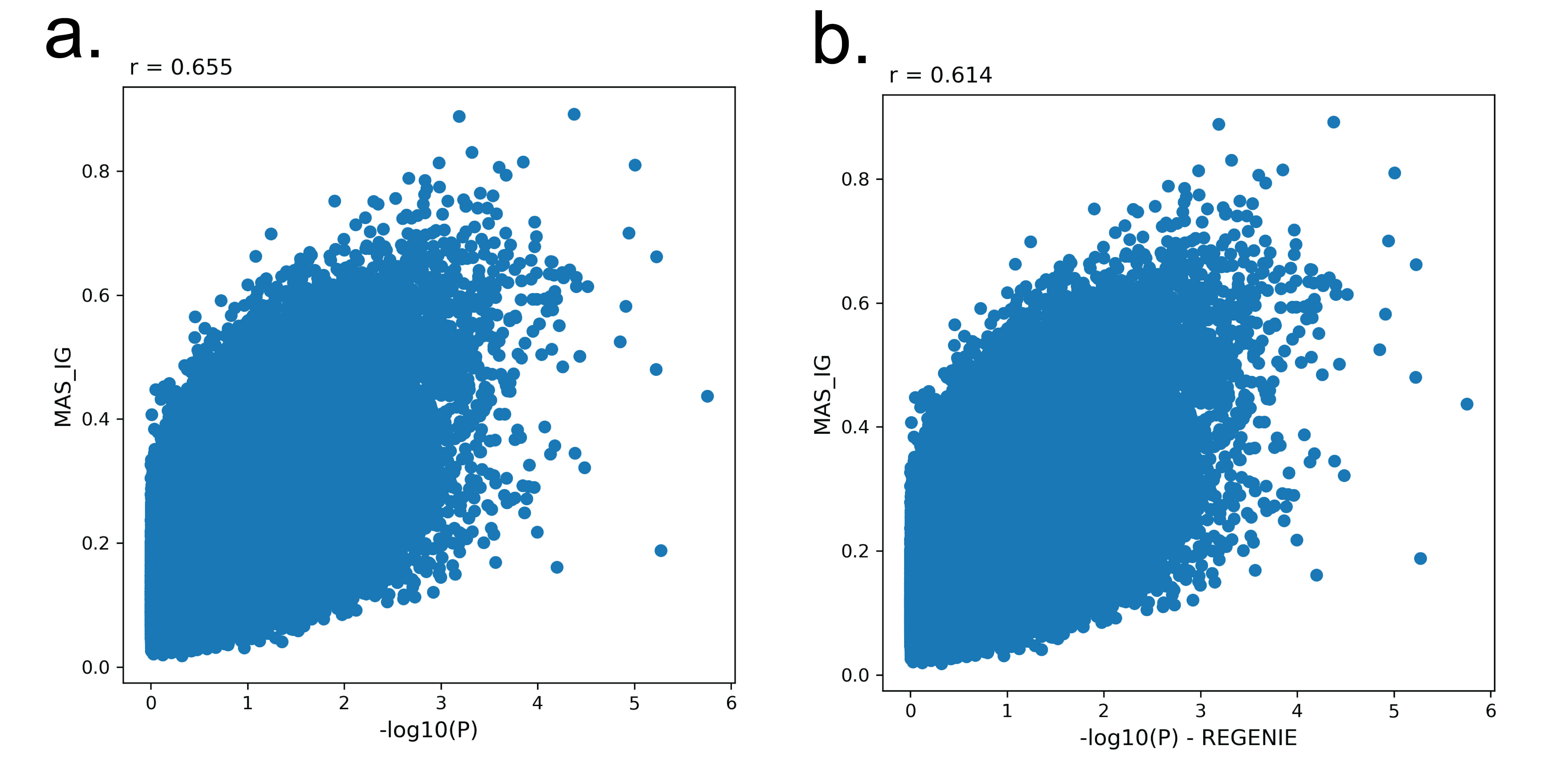}
\caption{Correlation between mean attribution score  (averaged over 10 different models) obtained via integrated gradient (MAS\_IG) approach and -log(P) values obtained via \textbf{a)} logistic regression and \textbf{b)} REGENIE association analysis using SCZ dataset. Provided r values are Pearson’s correlation coefficient.}
\label{suppfig:supfig5}
\end{figure*}

\begin{figure*}
\centering
\includegraphics[width=0.7\linewidth]{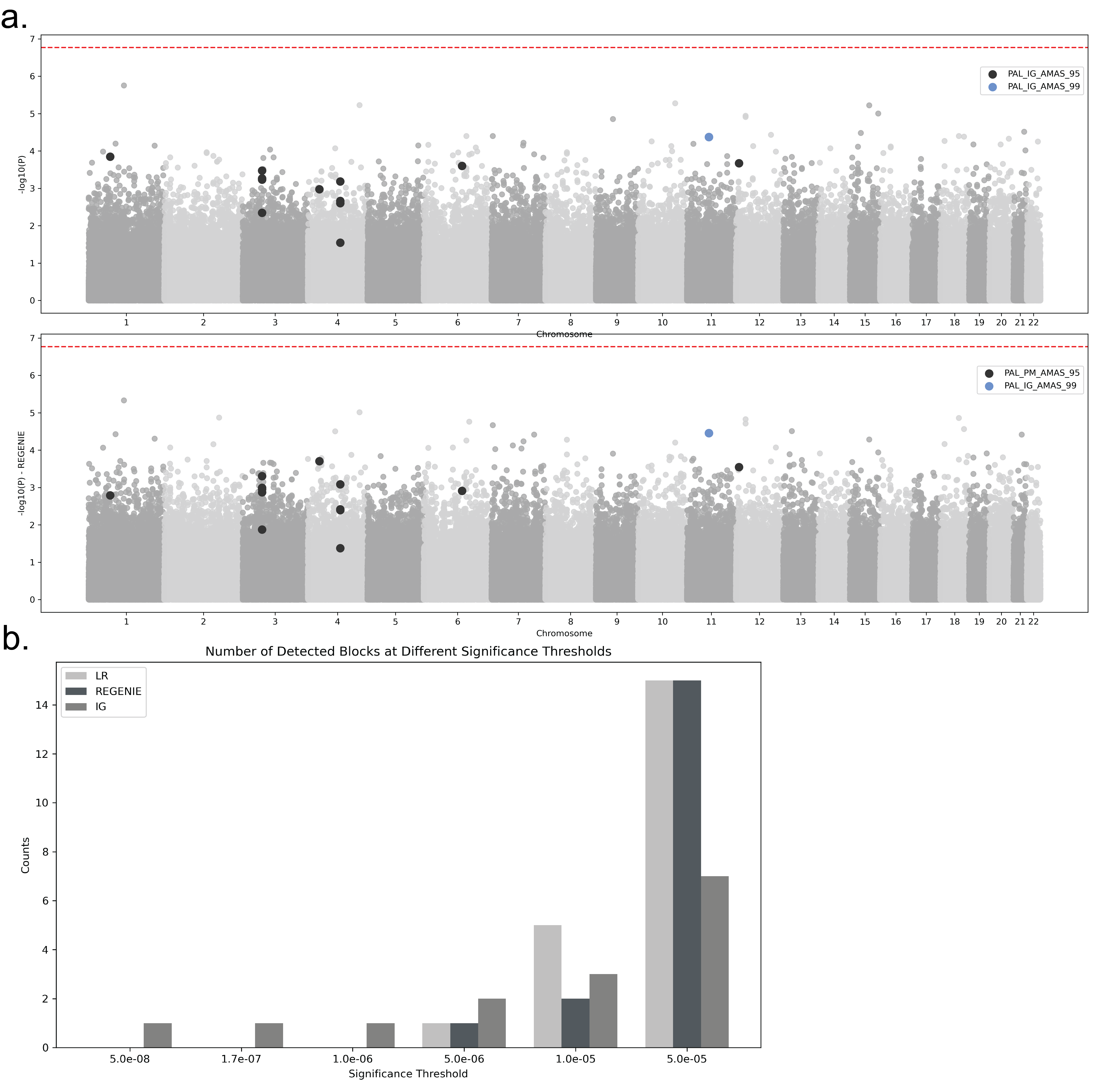}
\caption{\textbf{a)} Manhattan plots from LR (top) and REGENIE (bottom) association analysis with annotated $PAL_{AMAS}$ detected via IG approach for SCZ dataset. Blue and black dots represent $PAL_{AMAS}$ detected with strict (99.99 percentile) and relaxed (99.95 percentile) $\theta$ thresholds, respectively. Red dashed lines show Bonferroni significance thresholds for LR and REGENIE analyses. \textbf{b)} Number of detected blocks for IG, LR and REGENIE methods under different p-value significance thresholds. There was no common signal between IG and LR/REGENIE in any p-value threshold, whereas all signals detected by REGENIE were also detected by LR with 1e-05 and 5e-06 thresholds.}
\label{suppfig:supfig6}
\end{figure*}

\begin{figure*}
\centering
\includegraphics[width=0.7\linewidth]{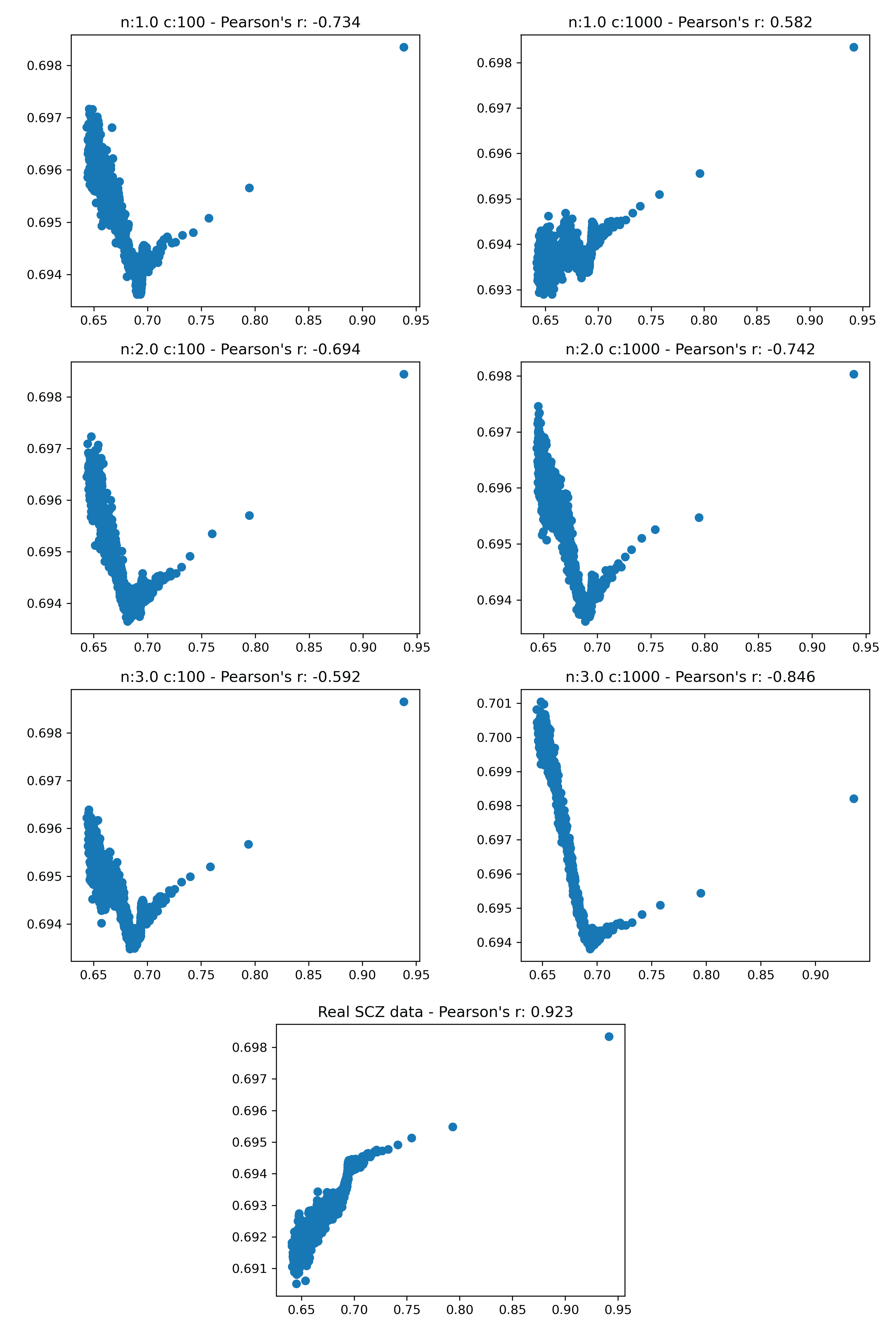}
\caption{Comparison of neural network model training using simulated and real data, based on the correlation between training loss (x-axis) and validation loss (y-axis). Training and validation losses were averaged over 10 different models. In each plot, the rightmost point corresponds to average loss after first epoch and leftmost point corresponds to average loss at the end of training (1000th epoch).}
\label{suppfig:supfig7}
\end{figure*}

\begin{figure*}
\centering
\includegraphics[width=1\linewidth]{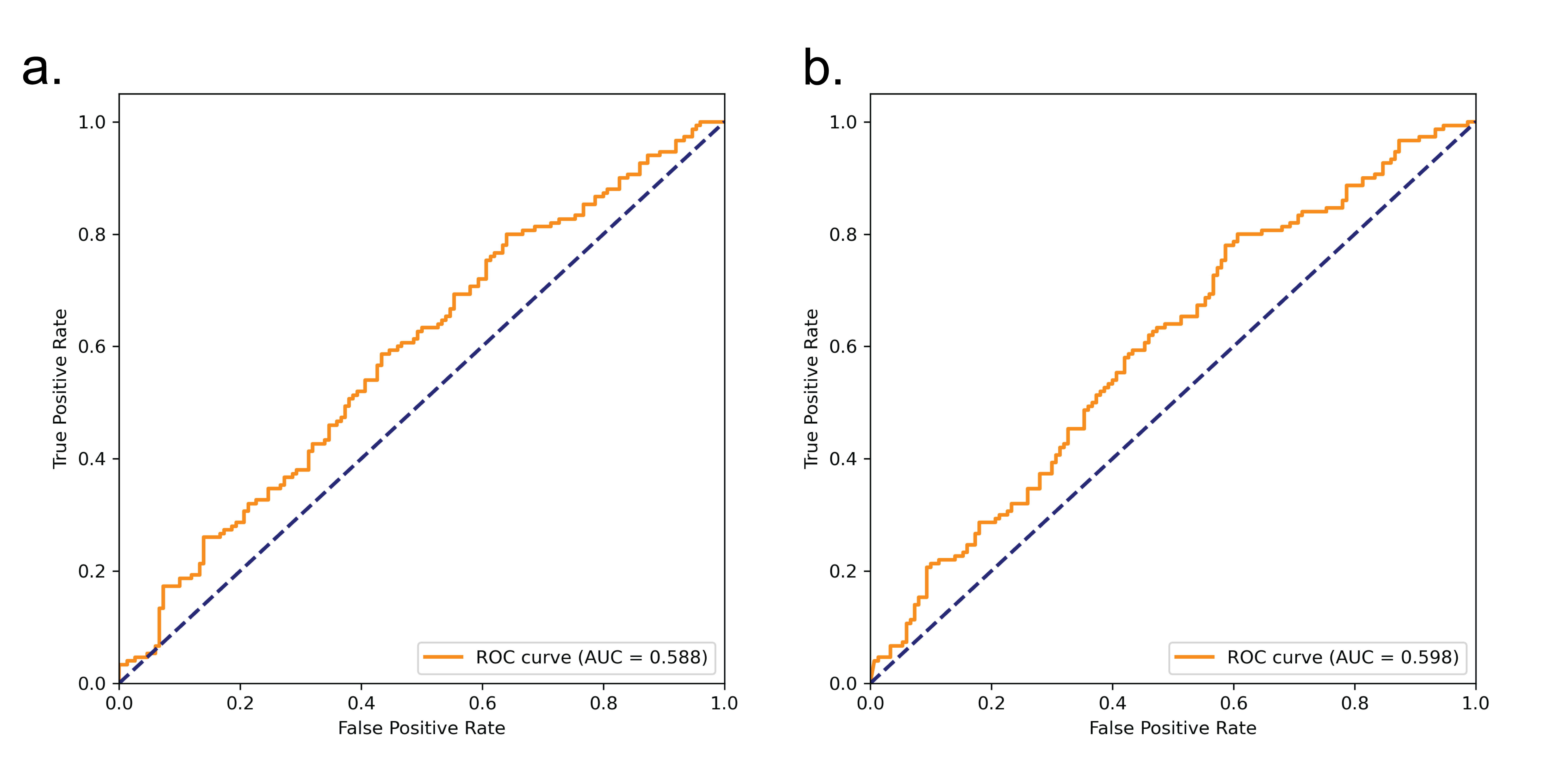}
\caption{Receiver operating characteristic (ROC) curves and area under the ROC curve (AUC) for \textbf{a)} neural network models trained for 1000 epochs (predictions averaged over 10 models trained with different seeds) and \textbf{b)} logistic regression phenotype predictions. Predictions were performed on the same test genotypes (150 cases, 150 controls) not utilised in fitting or training.}
\label{suppfig:supfig8}
\end{figure*}
